\documentclass{aa}  

\usepackage{graphicx}
\usepackage{txfonts}
\usepackage{placeins}

\usepackage[colorlinks=true,linkcolor=blue,citecolor=blue,filecolor=blue,urlcolor=blue]{hyperref}
\begin{document}

   \title{First step toward matter power spectrum reconstruction with Stage III weak gravitational lensing surveys}
   \titlerunning{Matter power spectrum reconstruction with Stage III WL surveys}
   \authorrunning{J. C. Broxterman and K. Kuijken}
   \author{Jeger C. Broxterman
          \inst{1,2}
          \and
          Konrad Kuijken \inst{1 }
          }

   \institute{Leiden Observatory, Leiden University, PO Box 9513, NL-2300 RA Leiden, The Netherlands\\
              \email{broxterman@strw.leidenuniv.nl}
         \and
             Lorentz Institute for Theoretical Physics, Leiden University, PO Box 9506, NL-2300 RA Leiden, the Netherlands\\
             }

   \date{Received September 20, 2024; accepted November 18, 2024}

 
  \abstract
   {Weak gravitational lensing (WL) surveys provide insight into the matter distribution over an extensive range of scales. Current WL results are in mild tension with cosmic microwave background measurements from the early Universe. Reconstructing the matter power spectrum from their measurements instead of condensing the information into a single cosmological parameter may help locate the origin of these differences.}
   {To investigate the cosmic shear measurements of Stage III WL surveys, we compared their tomographic data by assuming a simple parametric model for the matter power spectrum. The model allows the comparison of surveys with different characteristics and, in an agnostic approach, gives insight into the shape of the matter power spectrum preferred by the data without assuming a cosmological model.}
   {For the matter power spectrum, we assumed a double power-law model in scale factor and wavenumber. The best-fitting amplitude and exponents were inferred in a Markov chain Monte Carlo (MCMC) analysis. We identified the scales to which the data is most sensitive. We tested the sensitivity to different assumptions of the intrinsic alignment strength.}
   {We find that the constraining power of Stage III surveys on the power spectrum shape and evolution is still limited. Most information can be summarized as an overall amplitude at a pivot point in wavenumber and scale factor, while constraints on the power-law indices are considerably weaker. Nevertheless, all surveys show a weaker rate of growth from $z=$ 0.5 to 0.1 than predicted. The assumed intrinsic alignment strength is found to have no significant impact on the measured parameters and goodness of fit.}
   {Direct estimates of the matter power spectrum from Stage III weak lensing surveys can, in principle, be used to locate the physical origin of the observed $S_8$ tension. We present a simple methodology for the first steps in this direction, but find that current constraints are still weak.}


   \maketitle
%

\section{Introduction}
The concordance model of cosmology, the  Lambda-cold dark matter ($\Lambda$CDM) model, can fit an extensive range of observations with great accuracy \citep[see, e.g.,][]{Lahav2022}. Examples are the fluctuations in the cosmic microwave background \citep[CMB,][\textit{Planck}]{PlanckVI2020}; baryon acoustic oscillations \citep[BAOs;][]{DESI2024}; and cosmic shear, the slight distortion of galaxy images by weak gravitational lensing (WL) of the large-scale structure of the Universe \citep{Kilbinger2015}. With the increasing accuracy of cosmological surveys, several tensions between different probes have emerged. Most notably, the tension in the value of the Hubble constant ($H_\mathrm{0}$) inferred from high-redshift \textit{Planck} CMB measurements or local distance ladder measurements \citep{Riess2021}, and the value of $S_8 \equiv \sigma_\mathrm{8} \sqrt{\Omega_\mathrm{m}/0.3}$.\footnote{Where $\Omega_\mathrm{m}$ is the matter density parameter and $\sigma_8$ the amplitude of the linear theory matter power spectrum parameterized as the root mean squared mass density fluctuation in spheres of radius 8 $h^{-1}$ Mpc, both at $z = 0$.} Here the value of the \textit{Planck} CMB measurements of the early Universe is in slight tension with the value measured from cosmic shear measurement of the local Universe \citep{Hildebrandt2017,Asgari2021KiDS,Secco2022,Abbott2023,Li2023HSCyr3}. The main proposed solutions to solving the $S_\mathrm{8}$ tension involve unrecognized systematics or new physics \citep[see, e.g.,][]{Lucca2021,Tanimura2023,Joseph2023}.

The tension between the probes may have different physical origins. The CMB measures linear scales and the early Universe, whereas WL probes lower redshifts and  nonlinear scales. Assuming a physical origin, the tension may originate from differences in either scale (high vs. low wavenumber; $k$) or time (high vs. low redshift; $z$). 

\citet{Amon2022_Amod} explore the first option, introducing a one-parameter phenomenological model that modifies the power spectrum by fixing it at linear scales to the \textit{Planck} prediction and predicting the nonlinear suppression required to reconcile cosmic shear measurements from the Kilo-Degree Survey (KiDS) with \textit{Planck}. They remain agnostic as to what causes the suppression: obvious candidates are misinterpreted baryonic components or new physics. However, the model predicts the suppression of the nonlinear power spectrum using only dark matter (DM), and it does not contain any baryonic-informed components. Although baryonic suppression and nonlinear collapse manifest themselves in roughly the same wavenumber regime, the suppression they model is on slightly larger scales and has a different shape than predicted by state-of-the-art cosmological hydrodynamical simulation calibrated   primarily to X-ray data \citep[see, e.g., Fig.~1 of \citealt{Bigwood2024}, and, e.g.,][]{mcchartybahamas2017,Pakmor2023,Schaye2023,Schaller2024}.

\citet{Preston2023_DESyr3} extended the analysis to include the most recent cosmic shear measurement from the Dark Energy Survey (DES), and found that they require a somewhat lower suppression to resolve the tension with \textit{Planck}. However, the current measurements are not precise enough to constrain the baryonic feedback models. Recent findings suggest that the simulations calibrated to X-ray data, which probe the central regions of haloes, underpredict baryonic feedback as they cannot reproduce Sunyaev-Zel’dovich (SZ) measurements, which are more sensitive to the outskirts of haloes \citep{Amodeo2021,Bigwood2024,Hadzhiyska2024}. Although it might help resolve the $S_8$ tension, it raises a different problem because the X-ray and SZ measurements are no longer consistent with the same baryonic feedback model.

\begin{figure}
\centering
\includegraphics[width=\hsize]{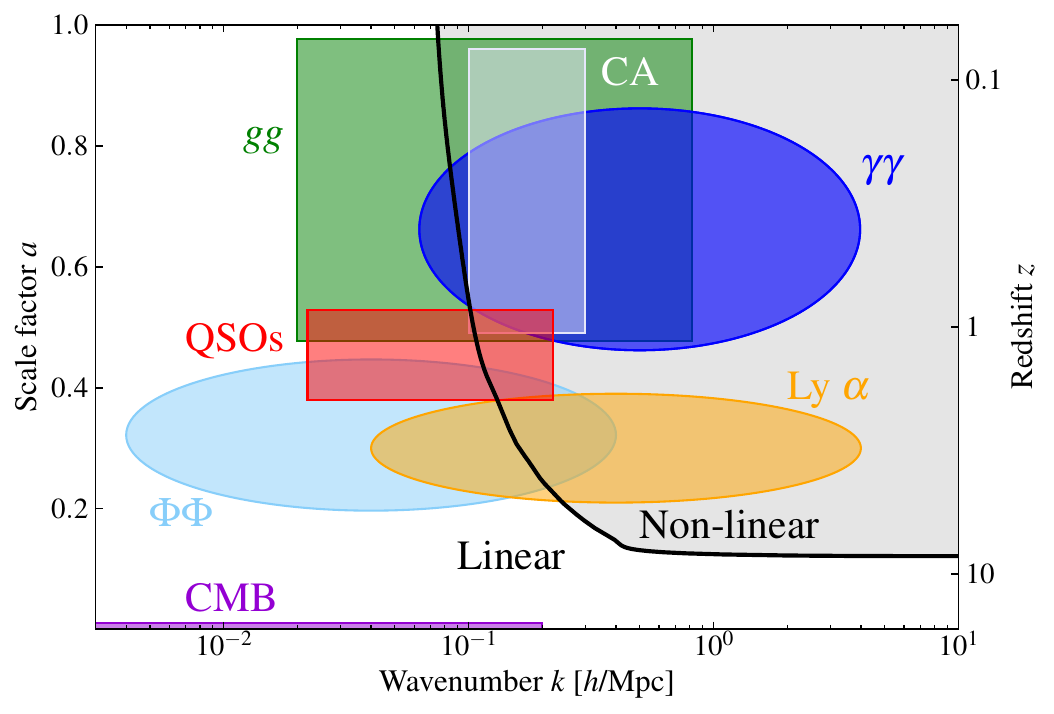}
  \caption{Observational cosmological probes of the matter power spectrum and the wavenumber ($k)$, scale factor ($a$), and redshift ($z$) ranges to which they are approximately sensitive. From early to late Universe: cosmic microwave background (CMB, purple), CMB lensing ($\Phi\Phi$, light blue), the Lyman-alpha forest (Ly $\alpha$, orange), quasi-stellar objects (QSOs, red), galaxy clustering ($gg$, green), cosmic shear ($\gamma\gamma$, dark blue), and cluster abundance (CA, white). The black curve separates the linear and nonlinear collapse regimes. Constraining the matter power spectrum from different probes   allows the localization of discrepancies in the two-dimensional ($k,a$) plane and provides insight into the physical origin of tensions between the measurements of different probes.}
     \label{fig:probe_sensitivity}
\end{figure}

In principle, we can distinguish tensions between measurements at different scales and times by directly constraining the matter power spectrum in the wavenumber ($k$) and scale factor ($a$) plane. Figure~\ref{fig:probe_sensitivity} illustrates the sensitivity of different observational cosmological probes in the ($k, a$) plane. The different probes are cosmic shear ($\gamma\gamma$), galaxy clustering \citep[$gg$; e.g.,][]{Reid2010,Gil-Marin2016}, CMB, CMB lensing \citep[$\Phi\Phi$; e.g.,][]{Planck2020CMBlensing}, cluster abundance \citep[CA; e.g.,][]{Tegmark2004}, quasi-stellar objects \citep[QSOs; e.g.,][]{DESI-III-2024}, and the Lyman alpha forest \citep[Ly $\alpha$; e.g.,][]{Palanque2015,Karacayli2024}. The squares correspond to probes where the emitted radiation is directly observed, whereas the ellipses correspond to processes happening along the line of sight between emission and observation. The black curve separates the linear and nonlinear collapse regimes. The estimate is obtained from \textsc{class} \citep[][]{Blas2011}{}{} by determining the wavenumber for which the nonlinear estimate differs by more than 5\% from the linear prediction. The difference between the different cosmic times and scales probed by the CMB and cosmic shear, as described above, is clearly visible in the figure. Comparing the different probes and their cross-correlations allows the localization of the tension, which should inform us regarding the origin of the observed discrepancies.

\citet{Preston2024_future} propose constraining the matter power spectrum from cosmic shear surveys. They estimate the constraining power of the matter power spectrum using synthetic cosmic shear measurements for the final data release of Stage III and Stage IV WL surveys. They find that while current cosmic shear surveys should only marginally be able to distinguish differences from a $\Lambda$CDM dark matter-only (DMO) model, next-generation surveys should provide clear constraints on the shape of the matter power spectrum at (mildly) nonlinear scales. 

Here, as a first step, we infer the best-fitting shape of the matter power spectrum using the latest cosmic shear measurement of Stage III WL surveys. We assume a simple parametric model that allows us to quantify the evolution of the matter power spectrum with $k$ and $a$ preferred by current data in an agnostic approach. The model allows the comparison of different surveys with different survey properties and does not assume a cosmological model for the matter power spectrum. We identify the scales to which the data is most sensitive, and test the impact of the assumed intrinsic alignment (IA) on the goodness of fit. 

In Sect.~\ref{sec:theory} we summarize the relevant WL theory to estimate the measured cosmic shear two-point statistics from the matter power spectrum. Section~\ref{sec:data} describes the data, and the assumed parametric model is introduced in Sect.~\ref{sec:methods}. In Sect.~\ref{sec:results} we present the measurements of the best-fitting parametric model of the matter power spectrum. We also compare our results to the best-fitting $\Lambda$CDM predictions. The main findings are summarized in Sect.~\ref{sec:conclusions}.


\section{Theory} \label{sec:theory}
In this section we briefly summarize the theoretical framework for cosmic shear measurements as projected statistics of the matter power spectrum. Cosmic shear measures the systematic distortion of galaxy images from WL by the large-scale structure. The measured signal is comprised of the signal induced by weak gravitational lensing (G), the intrinsic alignment of the galaxies (I), and the cross-correlation of the two effects. The observed cosmic shear angular power spectrum ($\mathcal{C}^{(ij)}_{\epsilon\epsilon}$) between two populations of galaxies in different tomographic bins $i$ and $j$, as a function of multipole moment $\ell$, is given by 
\begin{align}\label{eqn:powerspec}
    \mathcal{C}^{(ij)}_{\epsilon\epsilon}(\ell) =   \mathcal{C}^{(ij)}_{\mathrm{GG}}(\ell) + \mathcal{C}^{(ij)}_{\mathrm{GI}}(\ell)+\mathcal{C}^{(ij)}_{\mathrm{II}}(\ell).
\end{align}
Current cosmic shear surveys use $B$-mode statistics as null tests, and they find no evidence for nonzero $B$-modes or they limit their analysis to not include scales where the $B$-modes are nonzero due to unrecognized systematics. Here we assume that the $B$-modes are 0, and only use the $E$-mode signal such that $\mathcal{C}^{(ij)}_{\epsilon\epsilon}(\ell) = \mathcal{C}^{(ij)}_{\epsilon\epsilon,\mathrm{E}}(\ell)$. Assuming the modified Limber approximation, the angular power spectra are estimated from the matter power spectrum ($P_\mathrm{m}$) as \citep{Limber1953approx,Loverde2008limber}
\begin{align}\label{eqn:ang_power_spec}
    \mathcal{C}^{(ij)}_{\mathrm{ab}}(\ell) = \int_0^{\chi_{\mathrm{hor}}} \mathrm{d}\chi\, \frac{W^{(i)}_{\mathrm{a}}(\chi)W^{(j)}_{\mathrm{b}}(\chi)}{f^{2}_{\mathrm{K}}(\chi)}\, P_{\mathrm{m}}\bigg(\frac{\ell+1/2}{f_{\mathrm{K}}(\chi)},z(\chi)\bigg)\, ,
\end{align}
where $f_{\mathrm{K}}(\chi)$ is the angular diameter distance, $\chi$ the comoving distance, $z$ the redshift, and $\mathrm{a,b} \in {\mathrm{I},\mathrm{G}}$. The integral runs until the comoving line of sight horizon that is the edge of the galaxy survey ($\chi_{\mathrm{hor}}$). The WL kernel $W$ is given by \citep{Kaiser1992}
\begin{align}\label{eqn:G_kernal}
    W^{(i)}_{\mathrm{G}}(\chi) = \frac{3H_0^2\Omega_{\mathrm{m}}}{2c^2} \frac{f_{\mathrm{K}}(\chi)}{a(\chi)} \int_\chi^{\chi_{\mathrm{hor}}} \mathrm{d}\chi'\, n^{(i)}_{\mathrm{s}}(\chi')\, \frac{f_{\mathrm{K}}(\chi' - \chi)}{f_{\mathrm{K}}(\chi')},
\end{align}
where $c$ is the speed of light, $a = 1/(1+z)$ the scale factor, and $n_{\mathrm{s}}(\chi)$  the comoving distance source distribution that relates to the source redshift distribution as $n_\mathrm{s}(z)\mathrm{d}z = n_\mathrm{s}(\chi)\mathrm{d}\chi$. For the IA kernel, we assume the nonlinear alignment (NLA) model \citep{Bridle2007WIkernel}
\begin{align}\label{eqn:IA_kernal}
    W^{(i)}_{\mathrm{I}}(\chi) = -A_{\mathrm{IA}}\, \bigg(\frac{1+z(\chi)}{1+z_{\mathrm{piv}}}\bigg)^{\eta_{\mathrm{IA}}}\, \frac{C_1 \rho_{\mathrm{cr}} \Omega_{\mathrm{m}}}{D(a[\chi])} \,n^{(i)}(\chi),
\end{align}
with $A_\mathrm{IA}$ the IA amplitude; $z_\mathrm{piv}$ a finite redshift pivot point;   $\eta_{\mathrm{IA}}$ the redshift dependence exponent, which we set to 0;  $C_1 \rho_{\mathrm{cr}}\ = 0.0134$; and $D$ the linear growth factor. We assume a flat Universe such that $f_{\mathrm{K}}(\chi) = \chi$. We adopt either $A_{\mathrm{IA}} = 0$ or 1 to quantify whether including intrinsic alignment better fits the data. 

We used three different cosmic shear observables in our analysis. Each observable is a projection of the angular power spectra, but is sensitive to different scales as they depend on different filter functions. The first observable is the shear two-point correlation functions (2PCFs, $\xi_\mathrm{+/-}$):
\begin{align}\label{eqn:2PCFs}
    \xi_\mathrm{+/-}^{(ij)}(\theta) = \int_0^\infty \frac{\mathrm{d}\ell\ell}{2 \pi}\, \mathcal{J}_{0/4}(\ell \theta) \,\mathcal{C}^{(ij)}_{\epsilon\epsilon}(\ell).
\end{align}
Here $\mathcal{J}_{0}$ and $\mathcal{J}_{4}$ are cylindrical Bessel functions of the first kind of
the zeroth and fourth order, respectively. Second, we considered complete orthogonal sets of E/B-integrals (COSEBIs). An advantage of the COSEBIs is that nearly all cosmological information is contained in the first few $n$-modes, thus reducing the necessary computational time \citep{Asgari2012cosebis}. The COSEBIs ($E_n$) are estimated using \citep{Schneider2010cosebis}
\begin{align}\label{eqn:COSEBI_estimate}
    E_n^{(ij)} = \int_0^\infty \frac{\mathrm{d}\ell\ \ell}{2\pi}\, \mathcal{C}_{\epsilon\epsilon}^{(ij)}(\ell)\,W_n(\ell),
\end{align}
where the weight function, $W_n(\ell)$, is a Hankel transform given by Eq.~9 in \citet{Asgari2021KiDS}.  Similar to \citet{Asgari2021KiDS}, we used the first five COSEBIs $n$-modes.

Finally, we considered band power spectra ($\mathcal{C}_{\mathrm{E},l}$, band powers). This statistic is an angular average of the two-point correlation function that has a weak correlation between values at different wavenumbers and is given by \citep{Schneider2002} 
\begin{align}\label{eqn:bp_th}
    \mathcal{C}_{\mathrm{E},l}^{(ij)} &= \frac{1}{2\mathcal{N}_l} \int_{0}^{\infty} \mathrm{d}\ell\, W_{\mathrm{EE}}^l(\ell)\,\mathcal{C}_{\epsilon\epsilon}^{(ij)}(\ell),
\end{align}
where the normalisation $\mathcal{N}_l$ and kernels $W_{\mathrm{EE}}$ are respectively given by Eqs.~18 and 26 in \citet{Joachimi2021Kids}. We assume the best-fitting $\Lambda$CDM cosmology from \textit{Planck} TT,TE,EE+lowE+lensing for the values of the cosmological parameters and to compute the redshift-comoving distance relation and linear growth factor \citep{PlanckVI2020}. 

\section{Data}\label{sec:data}
We used the latest public data releases of the two-point statistics of three Stage III WL surveys: the Kilo-Degree Survey data release 4 \citep[KiDS-1000;][]{Kuijken2019KiDS1000}, the Dark Energy Survey year 3 \citep[DES year 3;][]{Abbott2022DESyr3}, and the Hyper Suprime-Cam Year 3 \citep[HSC year 3;][]{Dalal2023HSCyr3}. Collectively, these surveys are referred to as Stage III WL surveys. We use each survey's summary statistics, source redshift distributions, and full covariance matrix, as described in \citet{Asgari2021KiDS}, \citet{Hildebrandt2020}, and \citet{Joachimi2021Kids} for KiDS-1000; \citet{Abbott2022DESyr3}, \citet{Milesetal2021DESphotz}, and \citet{Friedrich2021covariance} for DES year 3; and \citet{Shirasaki19HSCCov}, \citet{Li2023HSCyr3}, and \citet{Rau2024HSCphotoz} for HSC year 3. We cut the data vectors at the same angular scales as the fiducial survey analyses. In Table~\ref{tab:surveys} we summarize the main characteristics of the surveys and the statistics we used in this analysis. The table lists the sky area covered, effective source number density ($n_{\mathrm{eff}}$), standard deviation of the galaxy shape measurement error, and number of tomographic bins. As indicated in the final column, for KiDS-1000 we used the 2PCFs, COSEBIs, and band powers, whereas for HSC and DES year 3 we only used the real space 2PCFs from \citet{Li2023HSCyr3} and \citet{Abbott2022DESyr3}, respectively. The former allows us to quantify the difference between the same measurements from a single survey and the scales that are probed by different statistics, whereas using all statistics allows us to quantify the variations between different surveys.

\begin{table*}
\centering
\caption{Survey characteristics.}
\label{tab:surveys}
\begin{tabular}{llllll}
\hline
Survey      & Area [deg$^2$] & $n_\mathrm{eff}$ [arcmin$^{-2}$]& $\sigma_e$ & No. of tomographic bins & statistics used in this analysis \\
\hline
KiDS-1000       & 1006 &  6.2 & 0.26 & 5 & $ \xi_\mathrm{+/-}$, $E_n$, $\mathcal{C}_{\mathrm{E},l}$ \\
DES year 3       & 4143 & 5.9 & 0.26 & 4 & $ \xi_\mathrm{+/-}$ \\
HSC year 3 & 416 & 15 & 0.24 & 4 & $ \xi_\mathrm{+/-}$ \\
\hline
\end{tabular}
\tablefoot{Listed for each survey:  sky area covered; total effective source number density, $n_{\mathrm{eff}}$;   standard deviation of the error of galaxy shape measurements, $\sigma_e$;   number of tomographic bins; and the statistics considered in this work.}
\end{table*}

\section{Methodology}\label{sec:methods}
Section~\ref{sec:theory}   provides the formalism for estimating the cosmic shear observables from the matter power spectrum. In this section we provide the model we used in our analysis. For the matter power spectrum, we assumed a double power law in wavenumber ($k$) and scale factor ($a$) as

\begin{align}\label{eq:double_power_law}
    P_\mathrm{m}(k,\,a) = A\, \bigg(\frac{k}{k_\mathrm{piv}}\bigg )^p\, \bigg(\frac{a}{a_\mathrm{piv}}\bigg)^m,
\end{align}
where the amplitude $A$, and exponents $p$ and $m$ are the parameters that will be fitted for. The pivot points $k_\mathrm{piv}$ and $a_\mathrm{piv}$ are free as they rescale the amplitude. We   used two sets of pivots. First, we determined the optimal values of the pivot points for each survey and statistic by finding the values of $k_\mathrm{piv}$ and $a_\mathrm{piv}$ for which the covariance between the amplitude $A$ and exponents $p$ and $m$ is zero. This method provides the power spectrum amplitude of the point on the ($k,a$) plane most tightly constrained by the statistic. We   provided the best-fitting parameters of the double power-law parameters assuming these pivot points. After, to more straightforwardly compare the three surveys, we adopted a common pivot for all surveys: $k_\mathrm{piv} = 0.5\, h/$Mpc and $a_\mathrm{piv} = 0.75$. We estimated the best-fitting parameters that maximize the log-likelihood by using the public Markov chain Monte Carlo (MCMC) sampler \textsc{emcee} \citep{Emcee2013}. We used flat priors in the range $\log_{\mathrm{10}} A \in [0,10],\, p \in [-2,0.5]$, and $m \in [-5,5]$. We computed the reduced chi-square value between the double power fits and measurements to quantify the goodness of fit.

\section{Results}\label{sec:results}
\subsection{Double power-law constraints}
We first list the results of determining the optimal pivot points for each survey. Table~\ref{tab:pivot_points} lists the best-fitting pivot points in wavenumber and scale factor for each of the five statistics considered in this work and for the choice of $A_{\mathrm{IA}} = 0$ or 1 in the third and fourth column, respectively. The fifth column gives the redshift corresponding to the scale factor pivot point ($z_\mathrm{piv} = 1/a_\mathrm{piv}-1$).

\begin{table*}
\centering
\caption{Pivot points and best-fitting parameters for the double power-law model.}
\label{tab:pivot_points}
\begin{tabular}{ll|lll|llll}
\hline
Statistic     & $A_\mathrm{IA}$ &  $k_{\mathrm{piv}}\, [h/$Mpc] & $a_{\mathrm{piv}}$ & $z_{\mathrm{piv}}$ & $\log_{10} A\, [\mathrm{Mpc}^3/h^3]$ & $p$ & $m$ & $\chi_\mathrm{red}^2$ \\
\hline
KiDS-1000 COSEBIs       & 0 & 1.20 &  0.75 & 0.34  & $2.04\pm 0.03 $ & $-1.30\pm0.04$ & $\phantom{-}0.3\pm0.8$ & 1.25\\
KiDS-1000 COSEBIs       & 1 & 1.24 & 0.75 & 0.34 & $2.03 \pm 0.03$  &  $-1.30\pm0.04$ & $\phantom{-}0.2\pm0.9$ & 1.26  \\
KiDS-1000 band powers       & 0  & 1.03 & 0.74 & 0.34 &  $2.16 \pm 0.03$ &  $-1.25\pm0.06$ & $-0.1\pm1.0$ & 1.39 \\
KiDS-1000 band powers       & 1 & 0.93 & 0.74& 0.35  &  $2.21 \pm 0.03$ &  $-1.26\pm0.06$ & $-0.1\pm1.1$ & 1.40\\
KiDS-1000 2PCFs       & 0 & 3.64 & 0.74& 0.35 &  $1.44\pm0.04$ & $-1.24\pm0.03$ & $\phantom{-}0.0\pm0.8$ & 1.25 \\
KiDS-1000 2PCFs       & 1 & 4.75 & 0.74& 0.35  &  $1.30 \pm 0.04$ &  $-1.24\pm0.03$ & $-0.3\pm0.8$ & 1.25\\
DES year 3 2PCFs      & 0 & 2.17 & 0.79& 0.26 & $1.92\pm0.03$ &  $-1.14\pm0.02$ & $\phantom{-}1.0\pm0.5$ & 1.14 \\
DES year 3 2PCFs      & 1 & 2.01 & 0.79& 0.26  & $1.97\pm0.03$ &  $-1.14\pm0.02$ & $\phantom{-}1.0\pm0.5$ & 1.14\\
HSC year 3 2PCFs & 0& 0.80 & 0.71& 0.41  & $2.36 \pm 0.03$ &  $-1.27 \pm 0.03 $ & $-0.2\pm0.4$ & 1.40\\
HSC year 3 2PCFs & 1 & 0.78 & 0.71&  0.41 & $2.28 \pm 0.03$ &  $-1.27\pm0.03$ & $-0.3\pm0.4$ & 1.38  \\
\hline
\end{tabular}
\tablefoot{Shown for each statistic and choice of intrinsic alignment amplitude ($A_\mathrm{IA}$) are the values of the pivot points that minimize the covariance between the amplitude $\log_{10} A$, and exponents $m$ and $p$ in wavenumber ($k_\mathrm{piv}$), scale factor ($a_\mathrm{piv}$), and corresponding redshift ($z_\mathrm{piv} = 1/a_\mathrm{piv}-1$) and the best-fitting parameters for the double power-law parameters; amplitude, $\log_{10} A$; wavenumber exponent, $p$; scale factor exponent, $m$. The final column indicates the reduced chi-square value of the double power-law fit, $\chi_\mathrm{red}^2$.}
\end{table*}

The values show that the HSC survey is the deepest, as the scale factor pivot point corresponds to $z \approx 0.4$. Then, the three KiDS-1000 statistics, which have almost identical depths, probe slightly less deep. Finally, the DES year 3 2PCFs probe the least deep.  The HSC 2PCFs probe the largest scales, whereas the KiDS-1000 2PCFs probe the smallest scales.

Table~\ref{tab:pivot_points} also includes the best-fitting parameter values and their 16th and 84th percentiles of the double power-law fit. For each fit the parameter best-fitting values correspond to an inference assuming the listed pivot point for that statistic and the value of intrinsic alignment amplitude. The best-fitting value of the amplitude parameter may therefore differ as this is the value of the double power-law model at ($k,a$) = $(k_\mathrm{piv},a_{\mathrm{piv}}$). In Appendix~\ref{app:cosebi_measurement} we show an example of a double-fit power-law prediction to the KiDS-1000 COSEBIs.

The three KiDS-1000 statistics generally find the same values for the power-law exponents $m$ and $p$, which are consistent with the HSC year 3 2PCFs estimates. The DES year 3 2PCFs show a weaker dependence on wavenumber and a stronger dependence on scale factor. 

Focussing first on the growth rate, quantified by the $m$ exponent, we see that the constraints from WL are weak. Most statistics are consistent with no growth, with only DES year 3 2PCFs preferring a positive growth rate of $m=1.0 \pm 0.5$ irrespective of the choice of $A_{\mathrm{IA}}$. 

Stage III WL surveys primarily probe the $\Lambda$-dominated era, during which linear growth eventually comes to a halt. However, this is a gradual process;  in addition, nonlinear growth continues for a longer time: predictions obtained with \textsc{class} \citep{Blas2011} indicate an increase in the power spectrum between redshift 0.5 and 0.1 of a factor of 1.5 (1.9) at wavenumber $k=0.1$ (1.0) $h/$Mpc, corresponding to $m$ values of 1.3 (2.1). Therefore, all the data sets we analyze here mildly favor a slower growth than predicted by $\Lambda$CDM. This is most apparent in the HSC year 3 2PCFs, which favor $m=-0.3 \pm 0.4$. 

Comparing the three different KiDS-1000 estimates for $p$, we find that the 2PCFs provide the tightest constraint. The $1\sigma$ uncertainty on $p$ is a factor of two smaller than the uncertainty provided by the band powers estimate. This is not unexpected: for the chosen configuration, the 2PCFs probe the largest range of scales, as illustrated in Fig.~1 of \citet{Asgari2021KiDS}.

Interestingly, the assumed value for $A_\mathrm{IA}$ does not markedly affect the results. There are no significant differences between the assumed values for $A_\mathrm{IA}$. The results using $A_\mathrm{IA} = 0$ or 1 are always consistent, and including the intrinsic alignment does not give a better fit, as indicated by the reduced chi-square value in the final column. The NLA model assumes that density perturbations are characterized by the Poisson equation. This assumption breaks down under nonlinear gravitational collapse, and therefore the model cannot physically capture the IA signal over the entire range of scales considered in this work \citep{Krause2016,Asgari2021KiDS}. However, as the current constraining power is still limited and the values we adopted for the IA strength correspond to those measured by current surveys, we expect that using a different model, such as the  tidal alignment and tidal torquing model \citep{Blazek2019}, or including a redshift \citep[e.g.,][]{Joachimi2011} or luminosity \citep[e.g.,][]{Chisari2016} dependence will not change the insensitivity to IA strength.  

The values of the reduced chi-square statistics are reasonable for the model's simplicity. For example, the model does not capture the BAO wiggles and can still fit the data over a broad range of  wavenumbers and redshifts. The reduced chi-square values for the KiDS-1000 COSEBIs are 1.25 ($A_\mathrm{IA} = 0$) and 1.26 ($A_\mathrm{IA} = 1$). These values indicate the fits are of equal quality to the best-fitting $\Lambda$CDM prediction from \citet{Asgari2021KiDS}, who quote a value of $\chi_\mathrm{red}^2 = 1.2$. The same holds for the KiDS-1000 band powers and 2PCFs.

\begin{figure*}
\centering
\includegraphics[width=\hsize]{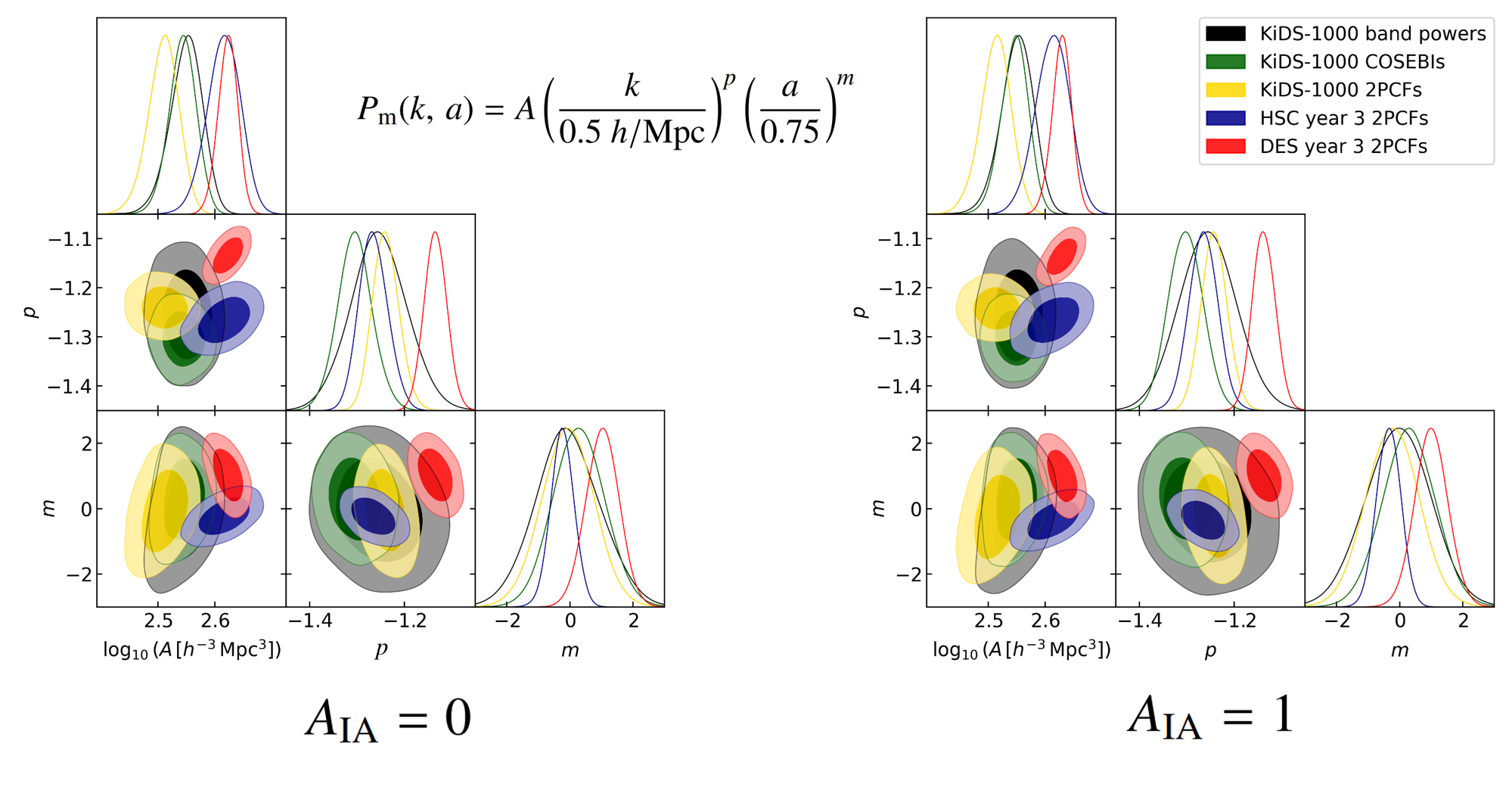}
  \caption{Posterior contours for a double power-law model as indicated at the top. The results correspond to setting the pivot points to $k_\mathrm{piv} = 0.5~h/$Mpc and $a_\mathrm{piv} = 0.75$. Left: Assuming no contribution from intrinsic alignment. Right: Contours from assuming an intrinsic alignment strength set by $A_\mathrm{IA} = 1$. The panels show the posterior distribution of the inferences using the KiDS-1000 band powers (black), COSEBIs (green), 2PCFs (yellow), HSC year 3 2PCFs (blue), and DES year 3 2PCFs (red). The results are consistent at the $\sim2\sigma$ level.}
     \label{fig:contours}
\end{figure*}

In Fig.~\ref{fig:contours} we show the best-fitting parameter constraints assuming a common pivot point for each survey. As the pivot point does not minimize the covariance between the different model parameters, clear correlations are visible between the amplitude parameter and $p$ and $m$ exponents. Even so, assuming the same pivot points allows for a direct comparison of the amplitudes measured by the different statistics at the same redshift and wavenumber. The left panel shows the results assuming an intrinsic alignment amplitude of 0, whereas the right panel shows   the contours obtained for $A_\mathrm{IA} = 1$. All KiDS-1000 and HSC year 3 results are consistent at the 2$\sigma$ level. Comparing the 1D amplitude $\log_{10}A$ posteriors, we see DES and HSC year 3 predict more power than the KiDS-1000 statistics at the $2\sigma$ level.

\subsection{Comparison to $\Lambda$CDM predictions}
\begin{figure}
\centering
\includegraphics[width=\hsize]{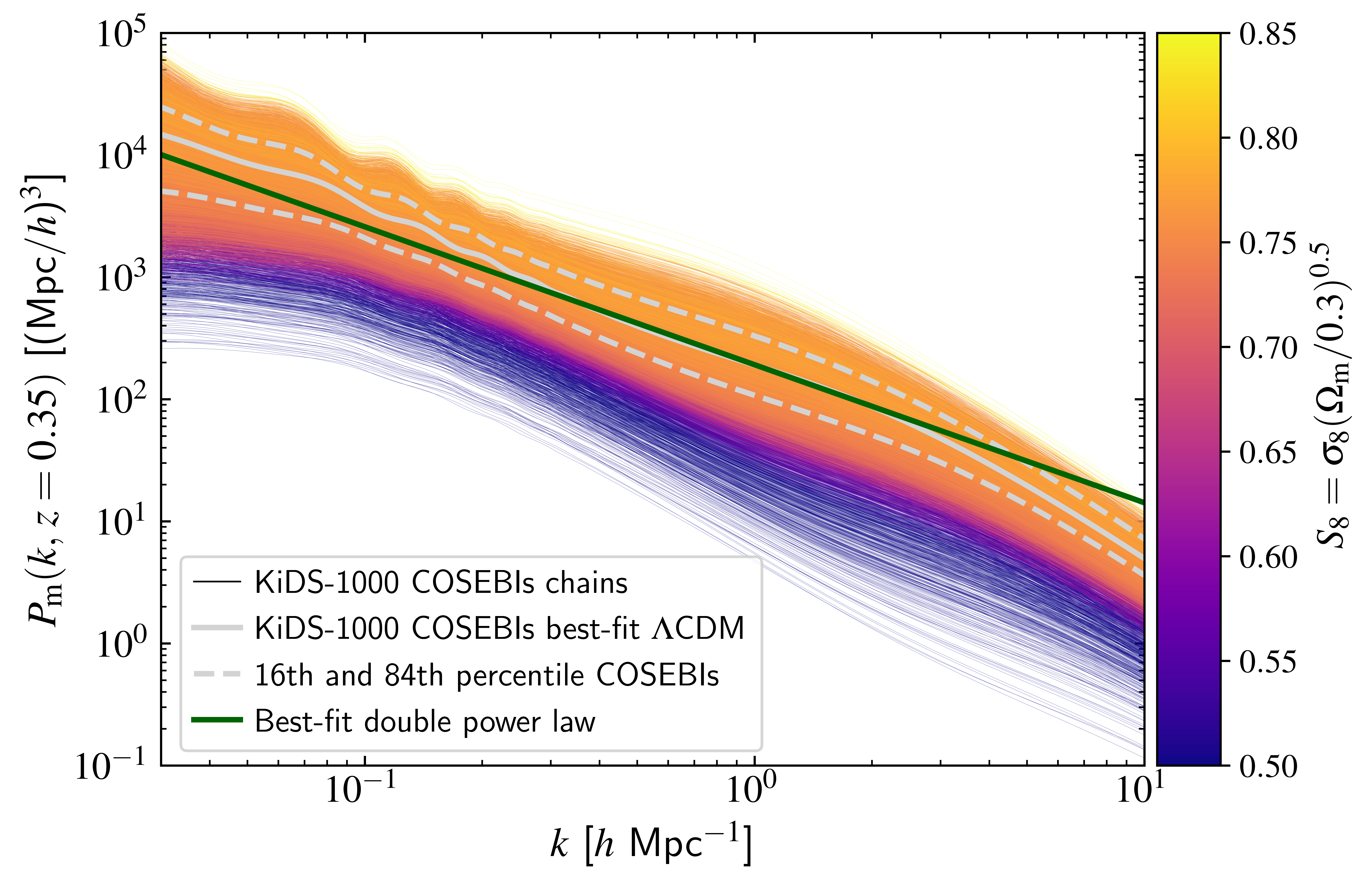}
  \caption{Matter power spectrum at $z=0.35$. The colored curves show the predictions from the KiDS-1000 COSEBIs chains from \citet{Asgari2021KiDS}, and they are color-coded by their value of $S_\mathrm{8}$ (see the color bar). The solid and dashed gray curves show the best-fitting and 16th and 84th percentiles of the KiDS-1000 inference, respectively. The green solid curve shows the best-fitting prediction of the $A_\mathrm{IA} = 1$ double power-law model assumed in this work.
          }
     \label{fig:pow_spec}
\end{figure}

Next, in Fig.~\ref{fig:pow_spec}, we compare our power-law prediction to the best-fitting $\Lambda$CDM predictions from the KiDS-1000 COSEBIs inference assuming $A_\mathrm{IA} = 1$. The green curve shows the double power prediction. The thin solid curves are all matter power spectrum predictions from the KiDS-1000 COSEBIs $\Lambda$CDM inference chains generated using \textsc{class} \citep{Blas2011} and \textsc{HMCode} \citep{Mead2015}. These curves are color-coded by their value of $S_8$, as indicated by the color bar. The solid and dashed gray curves respectively show the best-fitting and the 16th and 84th percentiles of the $\Lambda$CDM prediction from the KiDS-1000 COSEBIs \citep{Asgari2021KiDS}. The double power-law fit is consistent with the $\Lambda$CDM prediction over two decades in wavenumber. At the largest wavenumbers ($k > 0.5~h/$Mpc), the double power law overestimates the power compared to the $\Lambda$CDM prediction. The model also overpredicts the power on the largest scales ($k < 10^{-2}~h/$Mpc, not shown), which are beyond the scales to which cosmic shear is most sensitive. Within this range, the simplistic double power law is consistent with the KiDS-1000 COSEBIs $\Lambda$CDM prediction. The 16th and 84th percentile range typically covers half a dex, illustrating that the matter power spectrum is still poorly constrained.

\subsection{Suppression of the matter power spectrum}
\begin{figure}
\centering
\includegraphics[width=\hsize]{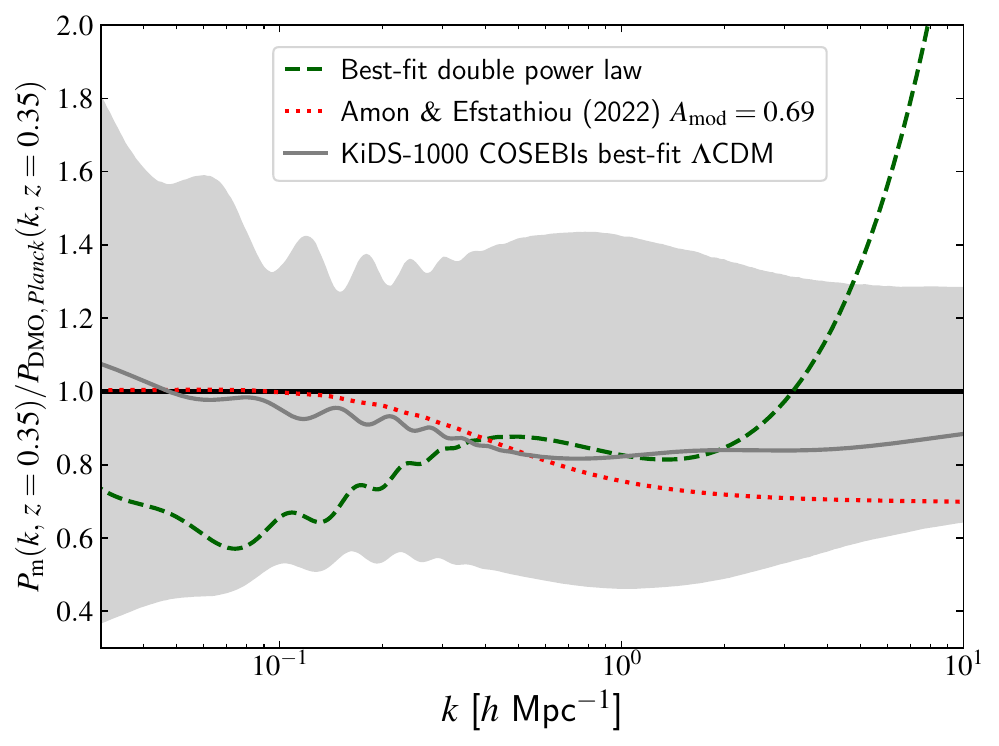}
  \caption{Suppression of the matter power spectrum compared to a DMO prediction at $z = 0.35$. The dashed green curve corresponds to the best-fitting prediction from the double power law model to the KiDS COSEBIs with $A_\mathrm{IA} = 1$. The solid gray curve shows the best fit of the KiDS-1000 $\Lambda$CDM inference from \citet{Asgari2021KiDS}, and the shaded area covers the 16th to 84th percentile region. The dotted red curve uses the parametric model from \cite{Amon2022_Amod} with $A_\mathrm{mod} = 0.69$.
    }
     \label{fig:ratios}
\end{figure}

Finally, we show the ratio of the matter power spectrum to a Planck DMO matter power spectrum as a function of wavenumber at $z=0.35$ in Fig.~\ref{fig:ratios}. The dashed green curve shows the best-fitting double power-law estimate from this work. The solid gray curves and shaded area respectively show the best-fitting and the  16th and 84th percentiles of the KiDS-1000 COSEBIs $\Lambda$CDM chains. The dotted red curve shows the prediction assuming the one-parameter parametric model from \citet{Amon2022_Amod} and setting $A_\mathrm{mod} = 0.69$, which is the value found by \citet{Amon2022_Amod} to reconcile KiDS-1000 with \textit{Planck}. 

The simple parametric model from \citet{Amon2022_Amod}, our double power law, and the DMO power spectrum are all within the 16th and 84th percentile prediction from the KiDS-1000 COSEBIs, which reconfirms that current data are not able to constrain detailed deviations from $\Lambda$CDM predictions or baryonic physics, as previously predicted in \citet{Preston2024_future}. 

\section{Conclusions}\label{sec:conclusions}
In this paper we have taken a first step toward reconstructing the shape of the matter power spectrum measured with Stage III WL lensing surveys. Reconstructing the matter power spectrum from different surveys that are sensitive to different redshift ranges and scales can give insights into the physical origins of observed tensions in cosmological parameters. We did not assume a cosmological model for the shape of the matter power spectrum, but instead remained agnostic to its shape by assuming a double power law in wavenumber and scale factor (Eq.~\ref{eq:double_power_law}). This allowed us to directly compare the results of different surveys and quantify the shape of the matter power spectrum preferred by the data. We identified the scales to which the data are most sensitive, and we tested different assumptions for the strength of the intrinsic alignment signal. The results highlight the potential of directly constraining the matter power spectrum to explore tensions in current data sets. We find the following:
   \begin{enumerate}
      \item Current cosmic shear surveys do not provide tight enough constraints on the shape of the matter power spectrum to rule out deviations from $\Lambda$CDM or realistic baryonic feedback models as the results are still consistent with DMO predictions.
      \item Stage III WL surveys can only poorly constrain the shape and evolution of the matter power spectrum in the low-redshift Universe. Even so, they favor a weaker evolution in the redshift range of 0.5 - 0.1 than predicted by $\Lambda$CDM.
      \item Including an intrinsic alignment of galaxies does not significantly impact the fitting accuracy or the best-fitting parameters.
   \end{enumerate}

More accurate measurements by next-generation WL surveys carried out by \textit{Euclid} \citep[][]{EuclidPaper1_2024}{}{}, \textit{Roman} \citep[][]{spergel2015wide}{}{}, and \textit{Rubin} \citep[][]{abell2009lsst}{}{} as well as more accurate modeling will help us  understand the physical origin of the tensions between different observational cosmological probes.


%
   \bibliographystyle{aa.bst} 
   \bibliography{references.bib} 

\begin{thebibliography}{56}
\expandafter\ifx\csname natexlab\endcsname\relax\def\natexlab#1{#1}\fi

\bibitem[{{Abbott} {et~al.}(2022){Abbott}, {Aguena}, {Alarcon}, {Allam},
  {Alves}, {Amon}, {Andrade-Oliveira}, {Annis}, {Avila}, {Bacon}, {Baxter},
  {Bechtol}, {Becker}, {Bernstein}, {Bhargava}, {Birrer}, {Blazek},
  {Brandao-Souza}, {Bridle}, {Brooks}, {Buckley-Geer}, {Burke}, {Camacho},
  {Campos}, {Carnero Rosell}, {Carrasco Kind}, {Carretero}, {Castander},
  {Cawthon}, {Chang}, {Chen}, {Chen}, {Choi}, {Conselice}, {Cordero},
  {Costanzi}, {Crocce}, {da Costa}, {da Silva Pereira}, {Davis}, {Davis}, {De
  Vicente}, {DeRose}, {Desai}, {Di Valentino}, {Diehl}, {Dietrich}, {Dodelson},
  {Doel}, {Doux}, {Drlica-Wagner}, {Eckert}, {Eifler}, {Elsner}, {Elvin-Poole},
  {Everett}, {Evrard}, {Fang}, {Farahi}, {Fernandez}, {Ferrero}, {Fert{\'e}},
  {Fosalba}, {Friedrich}, {Frieman}, {Garc{\'\i}a-Bellido}, {Gatti},
  {Gaztanaga}, {Gerdes}, {Giannantonio}, {Giannini}, {Gruen}, {Gruendl},
  {Gschwend}, {Gutierrez}, {Harrison}, {Hartley}, {Herner}, {Hinton},
  {Hollowood}, {Honscheid}, {Hoyle}, {Huff}, {Huterer}, {Jain}, {James},
  {Jarvis}, {Jeffrey}, {Jeltema}, {Kovacs}, {Krause}, {Kron}, {Kuehn},
  {Kuropatkin}, {Lahav}, {Leget}, {Lemos}, {Liddle}, {Lidman}, {Lima}, {Lin},
  {MacCrann}, {Maia}, {Marshall}, {Martini}, {McCullough}, {Melchior},
  {Mena-Fern{\'a}ndez}, {Menanteau}, {Miquel}, {Mohr}, {Morgan}, {Muir},
  {Myles}, {Nadathur}, {Navarro-Alsina}, {Nichol}, {Ogando}, {Omori},
  {Palmese}, {Pandey}, {Park}, {Paz-Chinch{\'o}n}, {Petravick}, {Pieres},
  {Plazas Malag{\'o}n}, {Porredon}, {Prat}, {Raveri}, {Rodriguez-Monroy},
  {Rollins}, {Romer}, {Roodman}, {Rosenfeld}, {Ross}, {Rykoff}, {Samuroff},
  {S{\'a}nchez}, {Sanchez}, {Sanchez}, {Sanchez Cid}, {Scarpine}, {Schubnell},
  {Scolnic}, {Secco}, {Serrano}, {Sevilla-Noarbe}, {Sheldon}, {Shin}, {Smith},
  {Soares-Santos}, {Suchyta}, {Swanson}, {Tabbutt}, {Tarle}, {Thomas}, {To},
  {Troja}, {Troxel}, {Tucker}, {Tutusaus}, {Varga}, {Walker}, {Weaverdyck},
  {Wechsler}, {Weller}, {Yanny}, {Yin}, {Zhang}, {Zuntz}, \& {DES
  Collaboration}}]{Abbott2022DESyr3}
{Abbott}, T.~M.~C., {Aguena}, M., {Alarcon}, A., {et~al.} 2022, \prd, 105,
  023520

\bibitem[{{Amodeo} {et~al.}(2021){Amodeo}, {Battaglia}, {Schaan}, {Ferraro},
  {Moser}, {Aiola}, {Austermann}, {Beall}, {Bean}, {Becker}, {Bond},
  {Calabrese}, {Calafut}, {Choi}, {Denison}, {Devlin}, {Duff}, {Duivenvoorden},
  {Dunkley}, {D{\"u}nner}, {Gallardo}, {Hall}, {Han}, {Hill}, {Hilton},
  {Hilton}, {Hlo{\v{z}}ek}, {Hubmayr}, {Huffenberger}, {Hughes}, {Koopman},
  {MacInnis}, {McMahon}, {Madhavacheril}, {Moodley}, {Mroczkowski}, {Naess},
  {Nati}, {Newburgh}, {Niemack}, {Page}, {Partridge}, {Schillaci}, {Sehgal},
  {Sif{\'o}n}, {Spergel}, {Staggs}, {Storer}, {Ullom}, {Vale}, {van Engelen},
  {Van Lanen}, {Vavagiakis}, {Wollack}, \& {Xu}}]{Amodeo2021}
{Amodeo}, S., {Battaglia}, N., {Schaan}, E., {et~al.} 2021, \prd, 103, 063514

\bibitem[{{Amon} \& {Efstathiou}(2022)}]{Amon2022_Amod}
{Amon}, A. \& {Efstathiou}, G. 2022, \mnras, 516, 5355

\bibitem[{{Asgari} {et~al.}(2021){Asgari}, {Lin}, {Joachimi}, {Giblin},
  {Heymans}, {Hildebrandt}, {Kannawadi}, {St{\"o}lzner}, {Tr{\"o}ster}, {van
  den Busch}, {Wright}, {Bilicki}, {Blake}, {de Jong}, {Dvornik}, {Erben},
  {Getman}, {Hoekstra}, {K{\"o}hlinger}, {Kuijken}, {Miller}, {Radovich},
  {Schneider}, {Shan}, \& {Valentijn}}]{Asgari2021KiDS}
{Asgari}, M., {Lin}, C.-A., {Joachimi}, B., {et~al.} 2021, \aap, 645, A104

\bibitem[{{Asgari} {et~al.}(2012){Asgari}, {Schneider}, \&
  {Simon}}]{Asgari2012cosebis}
{Asgari}, M., {Schneider}, P., \& {Simon}, P. 2012, \aap, 542, A122

\bibitem[{{Bigwood} {et~al.}(2024){Bigwood}, {Amon}, {Schneider}, {Salcido},
  {McCarthy}, {Preston}, {Sanchez}, {Sijacki}, {Schaan}, {Ferraro},
  {Battaglia}, {Chen}, {Dodelson}, {Roodman}, {Pieres}, {Fert{\'e}}, {Alarcon},
  {Drlica-Wagner}, {Choi}, {Navarro-Alsina}, {Campos}, {Ross}, {Carnero
  Rosell}, {Yin}, {Yanny}, {S{\'a}nchez}, {Chang}, {Davis}, {Doux}, {Gruen},
  {Rykoff}, {Huff}, {Sheldon}, {Tarsitano}, {Andrade-Oliveira}, {Bernstein},
  {Giannini}, {Diehl}, {Huang}, {Harrison}, {Sevilla-Noarbe}, {Tutusaus},
  {Elvin-Poole}, {McCullough}, {Zuntz}, {Blazek}, {DeRose}, {Cordero}, {Prat},
  {Myles}, {Eckert}, {Bechtol}, {Herner}, {Secco}, {Gatti}, {Raveri}, {Kind},
  {Becker}, {Troxel}, {Jarvis}, {MacCrann}, {Friedrich}, {Alves}, {Leget},
  {Chen}, {Rollins}, {Wechsler}, {Gruendl}, {Cawthon}, {Allam}, {Bridle},
  {Pandey}, {Everett}, {Shin}, {Hartley}, {Fang}, {Zhang}, {Aguena}, {Annis},
  {Bacon}, {Bertin}, {Bocquet}, {Brooks}, {Carretero}, {Castander}, {da Costa},
  {Pereira}, {De Vicente}, {Desai}, {Doel}, {Ferrero}, {Flaugher}, {Frieman},
  {Garc{\'\i}a-Bellido}, {Gaztanaga}, {Gutierrez}, {Hinton}, {Hollowood},
  {Honscheid}, {Huterer}, {James}, {Kuehn}, {Lahav}, {Lee}, {Marshall},
  {Mena-Fern{\'a}ndez}, {Miquel}, {Muir}, {Paterno}, {Plazas Malag{\'o}n},
  {Porredon}, {Romer}, {Samuroff}, {Sanchez}, {Sanchez Cid}, {Smith},
  {Soares-Santos}, {Suchyta}, {Swanson}, {Tarle}, {To}, {Weaverdyck}, {Weller},
  {Wiseman}, \& {Yamamoto}}]{Bigwood2024}
{Bigwood}, L., {Amon}, A., {Schneider}, A., {et~al.} 2024, \mnras, 534, 655

\bibitem[{{Blas} {et~al.}(2011){Blas}, {Lesgourgues}, \& {Tram}}]{Blas2011}
{Blas}, D., {Lesgourgues}, J., \& {Tram}, T. 2011, \jcap, 2011, 034

\bibitem[{{Blazek} {et~al.}(2019){Blazek}, {MacCrann}, {Troxel}, \&
  {Fang}}]{Blazek2019}
{Blazek}, J.~A., {MacCrann}, N., {Troxel}, M.~A., \& {Fang}, X. 2019, \prd,
  100, 103506

\bibitem[{{Bridle} \& {King}(2007)}]{Bridle2007WIkernel}
{Bridle}, S. \& {King}, L. 2007, New Journal of Physics, 9, 444

\bibitem[{{Chisari} {et~al.}(2016){Chisari}, {Laigle}, {Codis}, {Dubois},
  {Devriendt}, {Miller}, {Benabed}, {Slyz}, {Gavazzi}, \&
  {Pichon}}]{Chisari2016}
{Chisari}, N., {Laigle}, C., {Codis}, S., {et~al.} 2016, \mnras, 461, 2702

\bibitem[{{Dalal} {et~al.}(2023){Dalal}, {Li}, {Nicola}, {Zuntz}, {Strauss},
  {Sugiyama}, {Zhang}, {Rau}, {Mandelbaum}, {Takada}, {More}, {Miyatake},
  {Kannawadi}, {Shirasaki}, {Taniguchi}, {Takahashi}, {Osato}, {Hamana},
  {Oguri}, {Nishizawa}, {Malag{\'o}n}, {Sunayama}, {Alonso}, {Slosar}, {Luo},
  {Armstrong}, {Bosch}, {Hsieh}, {Komiyama}, {Lupton}, {Lust}, {MacArthur},
  {Miyazaki}, {Murayama}, {Nishimichi}, {Okura}, {Price}, {Tait}, {Tanaka}, \&
  {Wang}}]{Dalal2023HSCyr3}
{Dalal}, R., {Li}, X., {Nicola}, A., {et~al.} 2023, \prd, 108, 123519

\bibitem[{{Dark Energy Survey and Kilo-Degree Survey Collaboration}
  {et~al.}(2023){Dark Energy Survey and Kilo-Degree Survey Collaboration},
  {Abbott}, {Aguena}, {Alarcon}, {Alves}, {Amon}, {Andrade-Oliveira}, {Asgari},
  {Avila}, {Bacon}, {Bechtol}, {Becker}, {Bernstein}, {Bertin}, {Bilicki},
  {Blazek}, {Bocquet}, {Brooks}, {Burger}, {Burke}, {Camacho}, {Campos},
  {Carnero Rosell}, {Carrasco Kind}, {Carretero}, {Castander}, {Cawthon},
  {Chang}, {Chen}, {Choi}, {Conselice}, {Cordero}, {Crocce}, {da Costa}, {da
  Silva Pereira}, {Dalal}, {Davis}, {de Jong}, {DeRose}, {Desai}, {Diehl},
  {Dodelson}, {Doel}, {Doux}, {Drlica-Wagner}, {Dvornik}, {Eckert}, {Eifler},
  {Elvin-Poole}, {Everett}, {Fang}, {Ferrero}, {Fert{\'e}}, {Flaugher},
  {Friedrich}, {Frieman}, {Garc{\'\i}a-Bellido}, {Gatti}, {Giannini}, {Giblin},
  {Gruen}, {Gruendl}, {Gutierrez}, {Harrison}, {Hartley}, {Herner}, {Heymans},
  {Hildebrandt}, {Hinton}, {Hoekstra}, {Hollowood}, {Honscheid}, {Huang},
  {Huff}, {Huterer}, {James}, {Jarvis}, {Jeffrey}, {Jeltema}, {Joachimi},
  {Joudaki}, {Kannawadi}, {Krause}, {Kuehn}, {Kuijken}, {Kuropatkin}, {Lahav},
  {Leget}, {Lemos}, {Li}, {Li}, {Liddle}, {Lima}, {Lin}, {Lin}, {MacCrann},
  {Mahony}, {Marshall}, {McCullough}, {Mena-Fern{\'a}ndez}, {Menanteau},
  {Miquel}, {Mohr}, {Muir}, {Myles}, {Napolitano}, {Navarro-Alsina}, {Ogando},
  {Palmese}, {Pandey}, {Park}, {Paterno}, {Peacock}, {Petravick}, {Pieres},
  {Plazas Malag{\'o}n}, {Porredon}, {Prat}, {Radovich}, {Raveri}, {Reischke},
  {Robertson}, {Rollins}, {Romer}, {Roodman}, {Rykoff}, {Samuroff},
  {S{\'a}nchez}, {Sanchez}, {Sanchez}, {Schneider}, {Secco}, {Sevilla-Noarbe},
  {Shan}, {Sheldon}, {Shin}, {Sif{\'o}n}, {Smith}, {Soares-Santos},
  {St{\"o}lzner}, {Suchyta}, {Swanson}, {Tarle}, {Thomas}, {To}, {Troxel},
  {Tr{\"o}ster}, {Tutusaus}, {van den Busch}, {Varga}, {Walker}, {Weaverdyck},
  {Wechsler}, {Weller}, {Wiseman}, {Wright}, {Yanny}, {Yin}, {Yoon}, {Zhang},
  \& {Zuntz}}]{Abbott2023}
{Dark Energy Survey and Kilo-Degree Survey Collaboration}, {Abbott}, T.~M.~C.,
  {Aguena}, M., {et~al.} 2023, The Open Journal of Astrophysics, 6, 36

\bibitem[{{DESI Collaboration} {et~al.}(2024{\natexlab{a}}){DESI
  Collaboration}, {Adame}, {Aguilar}, {Ahlen}, {Alam}, {Alexander}, {Alvarez},
  {Alves}, {Anand}, {Andrade}, {Armengaud}, {Avila}, {Aviles}, {Awan},
  {Bahr-Kalus}, {Bailey}, {Baltay}, {Bault}, {Behera}, {BenZvi}, {Bera},
  {Beutler}, {Bianchi}, {Blake}, {Blum}, {Brieden}, {Brodzeller}, {Brooks},
  {Buckley-Geer}, {Burtin}, {Calderon}, {Canning}, {Carnero Rosell},
  {Cereskaite}, {Cervantes-Cota}, {Chabanier}, {Chaussidon}, {Chaves-Montero},
  {Chen}, {Chen}, {Claybaugh}, {Cole}, {Cuceu}, {Davis}, {Dawson}, {de la
  Macorra}, {de Mattia}, {Deiosso}, {Dey}, {Dey}, {Ding}, {Doel}, {Edelstein},
  {Eftekharzadeh}, {Eisenstein}, {Elliott}, {Fagrelius}, {Fanning}, {Ferraro},
  {Ereza}, {Findlay}, {Flaugher}, {Font-Ribera}, {Forero-S{\'a}nchez},
  {Forero-Romero}, {Frenk}, {Garcia-Quintero}, {Gazta{\~n}aga},
  {Gil-Mar{\'\i}n}, {Gontcho}, {Gonzalez-Morales}, {Gonzalez-Perez}, {Gordon},
  {Green}, {Gruen}, {Gsponer}, {Gutierrez}, {Guy}, {Hadzhiyska}, {Hahn},
  {Hanif}, {Herrera-Alcantar}, {Honscheid}, {Howlett}, {Huterer},
  {Ir{\v{s}}i{\v{c}}}, {Ishak}, {Juneau}, {Kara{\c{c}}ayl{\i}}, {Kehoe},
  {Kent}, {Kirkby}, {Kremin}, {Krolewski}, {Lai}, {Lan}, {Landriau}, {Lang},
  {Lasker}, {Le Goff}, {Le Guillou}, {Leauthaud}, {Levi}, {Li}, {Linder},
  {Lodha}, {Magneville}, {Manera}, {Margala}, {Martini}, {Maus}, {McDonald},
  {Medina-Varela}, {Meisner}, {Mena-Fern{\'a}ndez}, {Miquel}, {Moon}, {Moore},
  {Moustakas}, {Mudur}, {Mueller}, {Mu{\~n}oz-Guti{\'e}rrez}, {Myers},
  {Nadathur}, {Napolitano}, {Neveux}, {Newman}, {Nguyen}, {Nie}, {Niz},
  {Noriega}, {Padmanabhan}, {Paillas}, {Palanque-Delabrouille}, {Pan},
  {Penmetsa}, {Percival}, {Pieri}, {Pinon}, {Poppett}, {Porredon}, {Prada},
  {P{\'e}rez-Fern{\'a}ndez}, {P{\'e}rez-R{\`a}fols}, {Rabinowitz}, {Raichoor},
  {Ram{\'\i}rez-P{\'e}rez}, {Ramirez-Solano}, {Ravoux}, {Rashkovetskyi},
  {Rezaie}, {Rich}, {Rocher}, {Rockosi}, {Roe}, {Rosado-Marin}, {Ross},
  {Rossi}, {Ruggeri}, {Ruhlmann-Kleider}, {Samushia}, {Sanchez}, {Saulder},
  {Schlafly}, {Schlegel}, {Schubnell}, {Seo}, {Shafieloo}, {Sharples},
  {Silber}, {Slosar}, {Smith}, {Sprayberry}, {Tan}, {Tarl{\'e}}, {Taylor},
  {Trusov}, {Ure{\~n}a-L{\'o}pez}, {Vaisakh}, {Valcin}, {Valdes},
  {Vargas-Maga{\~n}a}, {Verde}, {Walther}, {Wang}, {Wang}, {Weaver},
  {Weaverdyck}, {Wechsler}, {Weinberg}, {White}, {Yu}, {Yu}, {Yuan},
  {Y{\`e}che}, {Zaborowski}, {Zarrouk}, {Zhang}, {Zhao}, {Zhao}, {Zhou},
  {Zhuang}, \& {Zou}}]{DESI2024}
{DESI Collaboration}, {Adame}, A.~G., {Aguilar}, J., {et~al.}
  2024{\natexlab{a}}, arXiv e-prints, arXiv:2404.03002

\bibitem[{{DESI Collaboration} {et~al.}(2024{\natexlab{b}}){DESI
  Collaboration}, {Adame}, {Aguilar}, {Ahlen}, {Alam}, {Alexander}, {Alvarez},
  {Alves}, {Anand}, {Andrade}, {Armengaud}, {Avila}, {Aviles}, {Awan},
  {Bailey}, {Baltay}, {Bault}, {Behera}, {BenZvi}, {Beutler}, {Bianchi},
  {Blake}, {Blum}, {Brieden}, {Brodzeller}, {Brooks}, {Buckley-Geer}, {Burtin},
  {Calderon}, {Canning}, {Carnero Rosell}, {Cereskaite}, {Cervantes-Cota},
  {Chabanier}, {Chaussidon}, {Chaves-Montero}, {Chen}, {Chen}, {Claybaugh},
  {Cole}, {Cuceu}, {Davis}, {Dawson}, {de la Macorra}, {de Mattia}, {Deiosso},
  {Dey}, {Dey}, {Ding}, {Doel}, {Edelstein}, {Eftekharzadeh}, {Eisenstein},
  {Elliott}, {Fagrelius}, {Fanning}, {Ferraro}, {Ereza}, {Findlay}, {Flaugher},
  {Font-Ribera}, {Forero-S{\'a}nchez}, {Forero-Romero}, {Garcia-Quintero},
  {Gazta{\~n}aga}, {Gil-Mar{\'\i}n}, {Gontcho}, {Gonzalez-Morales},
  {Gonzalez-Perez}, {Gordon}, {Green}, {Gruen}, {Gsponer}, {Gutierrez}, {Guy},
  {Hadzhiyska}, {Hahn}, {Hanif}, {Herrera-Alcantar}, {Honscheid}, {Howlett},
  {Huterer}, {Ir{\v{s}}i{\v{c}}}, {Ishak}, {Juneau}, {Kara{\c{c}}ayl{\i}},
  {Kehoe}, {Kent}, {Kirkby}, {Kremin}, {Krolewski}, {Lai}, {Lan}, {Landriau},
  {Lang}, {Lasker}, {Le Goff}, {Le Guillou}, {Leauthaud}, {Levi}, {Li},
  {Linder}, {Lodha}, {Magneville}, {Manera}, {Margala}, {Martini}, {Maus},
  {McDonald}, {Medina-Varela}, {Meisner}, {Mena-Fern{\'a}ndez}, {Miquel},
  {Moon}, {Moore}, {Moustakas}, {Mudur}, {Mueller}, {Mu{\~n}oz-Guti{\'e}rrez},
  {Myers}, {Nadathur}, {Napolitano}, {Neveux}, {Newman}, {Nguyen}, {Nie},
  {Niz}, {Noriega}, {Padmanabhan}, {Paillas}, {Palanque-Delabrouille}, {Pan},
  {Penmetsa}, {Percival}, {Pieri}, {Pinon}, {Poppett}, {Porredon}, {Prada},
  {P{\'e}rez-Fern{\'a}ndez}, {P{\'e}rez-R{\`a}fols}, {Rabinowitz}, {Raichoor},
  {Ram{\'\i}rez-P{\'e}rez}, {Ramirez-Solano}, {Rashkovetskyi}, {Rezaie},
  {Rich}, {Rocher}, {Rockosi}, {Roe}, {Rosado-Marin}, {Ross}, {Rossi},
  {Ruggeri}, {Ruhlmann-Kleider}, {Samushia}, {Sanchez}, {Saulder}, {Schlafly},
  {Schlegel}, {Schubnell}, {Seo}, {Sharples}, {Silber}, {Slosar}, {Smith},
  {Sprayberry}, {Swanson}, {Tan}, {Tarl{\'e}}, {Trusov}, {Vaisakh}, {Valcin},
  {Valdes}, {Vargas-Maga{\~n}a}, {Verde}, {Walther}, {Wang}, {Wang}, {Weaver},
  {Weaverdyck}, {Wechsler}, {Weinberg}, {White}, {Yu}, {Yu}, {Yuan},
  {Y{\`e}che}, {Zaborowski}, {Zarrouk}, {Zhang}, {Zhao}, {Zhao}, {Zhou}, \&
  {Zou}}]{DESI-III-2024}
{DESI Collaboration}, {Adame}, A.~G., {Aguilar}, J., {et~al.}
  2024{\natexlab{b}}, arXiv e-prints, arXiv:2404.03000

\bibitem[{{Euclid Collaboration} {et~al.}(2024){Euclid Collaboration},
  {Mellier}, {Abdurro'uf}, {Acevedo Barroso}, {Ach{\'u}carro}, {Adamek},
  {Adam}, {Addison}, {Aghanim}, {Aguena}, {Ajani}, {Akrami}, {Al-Bahlawan},
  {Alavi}, {Albuquerque}, {Alestas}, {Alguero}, {Allaoui}, {Allen}, {Allevato},
  {Alonso-Tetilla}, {Altieri}, {Alvarez-Candal}, {Alvi}, {Amara}, {Amendola},
  {Amiaux}, {Andika}, {Andreon}, {Andrews}, {Angora}, {Angulo}, {Annibali},
  {Anselmi}, {Anselmi}, {Arcari}, {Archidiacono}, {Aric{\`o}}, {Arnaud},
  {Arnouts}, {Asgari}, {Asorey}, {Atayde}, {Atek}, {Atrio-Barandela}, {Aubert},
  {Aubourg}, {Auphan}, {Auricchio}, {Aussel}, {Aussel}, {Avelino},
  {Avgoustidis}, {Avila}, {Awan}, {Azzollini}, {Baccigalupi}, {Bachelet},
  {Bacon}, {Baes}, {Bagley}, {Bahr-Kalus}, {Balaguera-Antolinez}, {Balbinot},
  {Balcells}, {Baldi}, {Baldry}, {Balestra}, {Ballardini}, {Ballester},
  {Balogh}, {Ba{\~n}ados}, {Barbier}, {Bardelli}, {Baron}, {Barreiro},
  {Barrena}, {Barriere}, {Barros}, {Barthelemy}, {Bartolo}, {Basset},
  {Battaglia}, {Battisti}, {Baugh}, {Baumont}, {Bazzanini}, {Beaulieu},
  {Beckmann}, {Belikov}, {Bel}, {Bellagamba}, {Bella}, {Bellini}, {Benabed},
  {Bender}, {Benevento}, {Bennett}, {Benson}, {Bergamini}, {Bermejo-Climent},
  {Bernardeau}, {Bertacca}, {Berthe}, {Berthier}, {Bethermin}, {Beutler},
  {Bevillon}, {Bhargava}, {Bhatawdekar}, {Bianchi}, {Bisigello}, {Biviano},
  {Blake}, {Blanchard}, {Blazek}, {Blot}, {Bosco}, {Bodendorf}, {Boenke},
  {B{\"o}hringer}, {Boldrini}, {Bolzonella}, {Bonchi}, {Bonici}, {Bonino},
  {Bonino}, {Bonvin}, {Bon}, {Booth}, {Borgani}, {Borlaff}, {Borsato}, {Bosco},
  {Bose}, {Botticella}, {Boucaud}, {Bouche}, {Boucher}, {Boutigny}, {Bouvard},
  {Bouwens}, {Bouy}, {Bowler}, {Bozza}, {Bozzo}, {Branchini}, {Brando},
  {Brau-Nogue}, {Brekke}, {Bremer}, {Brescia}, {Breton}, {Brinchmann},
  {Brinckmann}, {Brockley-Blatt}, {Brodwin}, {Brouard}, {Brown}, {Bruton},
  {Bucko}, {Buddelmeijer}, {Buenadicha}, {Buitrago}, {Burger}, {Burigana},
  {Busillo}, {Busonero}, {Cabanac}, {Cabayol-Garcia}, {Cagliari}, {Caillat},
  {Caillat}, {Calabrese}, {Calabro}, {Calderone}, {Calura}, {Camacho Quevedo},
  {Camera}, {Campos}, {Canas-Herrera}, {Candini}, {Cantiello}, {Capobianco},
  {Cappellaro}, {Cappelluti}, {Cappi}, {Caputi}, {Cara}, {Carbone}, {Cardone},
  {Carella}, {Carlberg}, {Carle}, {Carminati}, {Caro}, {Carrasco}, {Carretero},
  {Carrilho}, {Carron Duque}, {Carry}, {Carvalho}, {Carvalho}, {Casas},
  {Casas}, {Casenove}, {Casey}, {Cassata}, {Castander}, {Castelao},
  {Castellano}, {Castiblanco}, {Castignani}, {Castro}, {Cavet}, {Cavuoti},
  {Chabaud}, {Chambers}, {Charles}, {Charlot}, {Chartab}, {Chary}, {Chaumeil},
  {Cho}, {Chon}, {Ciancetta}, {Ciliegi}, {Cimatti}, {Cimino}, {Cioni},
  {Claydon}, {Cleland}, {Cl{\'e}ment}, {Clements}, {Clerc}, {Clesse}, {Codis},
  {Cogato}, {Colbert}, {Cole}, {Coles}, {Collett}, {Collins}, {Colodro-Conde},
  {Colombo}, {Combes}, {Conforti}, {Congedo}, {Conseil}, {Conselice},
  {Contarini}, {Contini}, {Conversi}, {Cooray}, {Copin}, {Corasaniti},
  {Corcho-Caballero}, {Corcione}, {Cordes}, {Corpace}, {Correnti}, {Costanzi},
  {Costille}, {Courbin}, {Courcoult Mifsud}, {Courtois}, {Cousinou}, {Covone},
  {Cowell}, {Cragg}, {Cresci}, {Cristiani}, {Crocce}, {Cropper}, {E Crouzet},
  {Csizi}, {Cuby}, {Cucchetti}, {Cucciati}, {Cuillandre}, {Cunha}, {Cuozzo},
  {Daddi}, {D'Addona}, {Dafonte}, {Dagoneau}, {Dalessandro}, {Dalton},
  {D'Amico}, {Dannerbauer}, {Danto}, {Das}, {Da Silva}, {da Silva},
  {d'Assignies Doumerg}, {Daste}, {Davies}, {Davini}, {Dayal}, {de Boer},
  {Decarli}, {De Caro}, {Degaudenzi}, {Degni}, {de Jong}, {de la Bella}, {de la
  Torre}, {Delhaise}, {Delley}, {Delucchi}, {De Lucia}, {Denniston}, {De
  Paolis}, {De Petris}, {Derosa}, {Desai}, {Desjacques}, {Despali}, {Desprez},
  {De Vicente-Albendea}, {Deville}, {Dias}, {D{\'\i}az-S{\'a}nchez}, {Diaz},
  {Di Domizio}, {Diego}, {Di Ferdinando}, {Di Giorgio}, {Dimauro}, {Dinis},
  {Dolag}, {Dolding}, {Dole}, {Dom{\'\i}nguez S{\'a}nchez}, {Dor{\'e}},
  {Dournac}, {Douspis}, {Dreihahn}, {Droge}, {Dryer}, {Dubath}, {Duc},
  {Ducret}, {Duffy}, {Dufresne}, {Duncan}, {Dupac}, {Duret}, {Durrer},
  {Durret}, {Dusini}, {Ealet}, {Eggemeier}, {Eisenhardt}, {Elbaz}, {Elkhashab},
  {Ellien}, {Endicott}, {Enia}, {Erben}, {Escartin Vigo}, {Escoffier},
  {Escudero Sanz}, {Essert}, {Ettori}, {Ezziati}, {Fabbian}, {Fabricius},
  {Fang}, {Farina}, {Farina}, {Farinelli}, {Farrens}, {Faustini}, {Feltre},
  {Ferguson}, {Ferrando}, {Ferrari}, {Ferr{\'e}-Mateu}, {Ferreira}, {Ferreras},
  {Ferrero}, {Ferriol}, {Ferruit}, {Filleul}, {Finelli}, {Finkelstein},
  {Finoguenov}, {Fiorini}, {Flentge}, {Focardi}, {Fonseca}, {Fontana},
  {Fontanot}, {Fornari}, {Fosalba}, {Fossati}, {Fotopoulou}, {Fouchez},
  {Fourmanoit}, {Frailis}, {Fraix-Burnet}, {Franceschi}, {Franco}, {Franzetti},
  {Freihoefer}, {Frenk}, {Frittoli}, {Frugier}, {Frusciante}, {Fumagalli},
  {Fumagalli}, {Fumana}, {Fu}, {Gabarra}, {Galeotta}, {Galluccio}, {Ganga},
  {Gao}, {Garc{\'\i}a-Bellido}, {Garcia}, {Gardner}, {Garilli},
  {Gaspar-Venancio}, {Gasparetto}, {Gautard}, {Gavazzi}, {Gaztanaga},
  {Genolet}, {Genova Santos}, {Gentile}, {George}, {Gerbino}, {Ghaffari},
  {Giacomini}, {Gianotti}, {Gibb}, {Gillard}, {Gillis}, {Ginolfi}, {Giocoli},
  {Girardi}, {Giri}, {Goh}, {G{\'o}mez-Alvarez}, {Gonzalez-Perez}, {Gonzalez},
  {Gonzalez}, {Gonzalez}, {Gouyou Beauchamps}, {Gozaliasl}, {Gracia-Carpio},
  {Grandis}, {Granett}, {Granvik}, {Grazian}, {Gregorio}, {Grenet}, {Grillo},
  {Grupp}, {Gruppioni}, {Gruppuso}, {Guerbuez}, {Guerrini}, {Guidi},
  {Guillard}, {Gutierrez}, {Guttridge}, {Guzzo}, {Gwyn}, {Haapala}, {Haase},
  {Haddow}, {Hailey}, {Hall}, {Hall}, {Hamaus}, {Haridasu},
  {Harnois-D{\'e}raps}, {Harper}, {Hartley}, {Hasinger}, {Hassani}, {Hatch},
  {Haugan}, {H{\"a}u{\ss}ler}, {Heavens}, {Heisenberg}, {Helmi}, {Helou},
  {Hemmati}, {Henares}, {Herent}, {Hern{\'a}ndez-Monteagudo}, {Heuberger},
  {Hewett}, {Heydenreich}, {Hildebrandt}, {Hirschmann}, {Hjorth}, {Hoar},
  {Hoekstra}, {Holland}, {Holliman}, {Holmes}, {Hook}, {Horeau}, {Hormuth},
  {Hornstrup}, {Hosseini}, {Hu}, {Hudelot}, {Hudson}, {Huertas-Company},
  {Huff}, {Hughes}, {Humphrey}, {Hunt}, {Huynh}, {Ibata}, {Ichikawa},
  {Iglesias-Groth}, {Ilbert}, {Ili{\'c}}, {Ingoglia}, {Iodice}, {Israel},
  {Israelsson}, {Izzo}, {Jablonka}, {Jackson}, {Jacobson}, {Jafariyazani},
  {Jahnke}, {Jain}, {Jansen}, {Jarvis}, {Jasche}, {Jauzac}, {Jeffrey},
  {Jhabvala}, {Jimenez-Teja}, {Jimenez Mu{\~n}oz}, {Joachimi}, {Johansson},
  {Joudaki}, {Jullo}, {Kajava}, {Kang}, {Kannawadi}, {Kansal}, {Karagiannis},
  {K{\"a}rcher}, {Kashlinsky}, {Kazandjian}, {Keck}, {Keih{\"a}nen}, {Kerins},
  {Kermiche}, {Khalil}, {Kiessling}, {Kiiveri}, {Kilbinger}, {Kim}, {King},
  {Kirkpatrick}, {Kitching}, {Kluge}, {Knabenhans}, {Knapen}, {Knebe}, {Kneib},
  {Kohley}, {Koopmans}, {Koskinen}, {Koulouridis}, {Kou}, {Kov{\'a}cs},
  {Kova{\v{c}}i{\'c}}, {Kowalczyk}, {Koyama}, {Kraljic}, {Krause}, {Kruk},
  {Kubik}, {Kuchner}, {Kuijken}, {K{\"u}mmel}, {Kunz}, {Kurki-Suonio},
  {Lacasa}, {Lacey}, {La Franca}, {Lagarde}, {Lahav}, {Laigle}, {La Marca}, {La
  Marle}, {Lamine}, {Lam}, {Lan{\c{c}}on}, {Landt}, {Langer}, {Lapi},
  {Larcheveque}, {Larsen}, {Lattanzi}, {Laudisio}, {Laugier}, {Laureijs},
  {Laurent}, {Lavaux}, {Lawrenson}, {Lazanu}, {Lazeyras}, {Le Boulc'h}, {Le
  Brun}, {Le Brun}, {Leclercq}, {Lee}, {Le Graet}, {Legrand}, {Leirvik}, {Le
  Jeune}, {Lembo}, {Le Mignant}, {Lepinzan}, {Lepori}, {Le Reun}, {Leroy},
  {Lesci}, {Lesgourgues}, {Leuzzi}, {Levi}, {Liaudat}, {Libet}, {Liebing},
  {Ligori}, {Lilje}, {Lin}, {Linde}, {Linder}, {Lindholm}, {Linke}, {Li},
  {Liu}, {Lloro}, {Lobo}, {Lodieu}, {Lombardi}, {Lombriser}, {Lonare}, {Longo},
  {L{\'o}pez-Caniego}, {Lopez Lopez}, {Alvarez}, {Loureiro}, {Loveday},
  {Lusso}, {Macias-Perez}, {Maciaszek}, {Maggio}, {Magliocchetti}, {Magnard},
  {Magnier}, {Magro}, {Mahler}, {Mainetti}, {Maino}, {Maiorano}, {Maiorano},
  {Malavasi}, {Mamon}, {Mancini}, {Mandelbaum}, {Manera},
  {Manj{\'o}n-Garc{\'\i}a}, {Mannucci}, {Mansutti}, {Manteiga Outeiro},
  {Maoli}, {Maraston}, {Marcin}, {Marcos-Arenal}, {Margalef-Bentabol},
  {Marggraf}, {Marinucci}, {Marinucci}, {Markovic}, {Marleau}, {Marpaud},
  {Martignac}, {Mart{\'\i}n-Fleitas}, {Martin-Moruno}, {Martin}, {Martinelli},
  {Martinet}, {Martin}, {Martins}, {Marulli}, {Massari}, {Massey}, {Masters},
  {Matarrese}, {Matsuoka}, {Matthew}, {Maughan}, {Mauri}, {Maurin},
  {Maurogordato}, {McCarthy}, {McConnachie}, {McCracken}, {McDonald}, {McEwen},
  {McPartland}, {Medinaceli}, {Mehta}, {Mei}, {Melchior}, {Melin},
  {M{\'e}nard}, {Mendes}, {Mendez-Abreu}, {Meneghetti}, {Mercurio}, {Merlin},
  {Metcalf}, {Meylan}, {Migliaccio}, {Mignoli}, {Miller}, {Miluzio},
  {Milvang-Jensen}, {Mimoso}, {Miquel}, {Miyatake}, {Mobasher}, {Mohr},
  {Monaco}, {Mongui{\'o}}, {Montoro}, {Mora}, {Moradinezhad Dizgah}, {Moresco},
  {Moretti}, {Morgante}, {Morisset}, {Moriya}, {Morris}, {Mortlock},
  {Moscardini}, {Mota}, {Mottet}, {Moustakas}, {Moutard}, {M{\"u}ller},
  {Munari}, {Murphree}, {Murray}, {Murray}, {Musi}, {Nadathur}, {Nagam},
  {Nagao}, {Naidoo}, {Nakajima}, {Nally}, {Natoli}, {Navarro-Alsina}, {Navarro
  Girones}, {Neissner}, {Nersesian}, {Nesseris}, {Nguyen-Kim}, {Nicastro},
  {Nichol}, {Nielbock}, {Niemi}, {Nieto}, {Nilsson}, {Noller}, {Norberg},
  {Nouri-Zonoz}, {Ntelis}, {Nucita}, {Nugent}, {Nunes}, {Nutma}, {Ocampo},
  {Odier}, {Oesch}, {Oguri}, {Magalhaes Oliveira}, {Onoue}, {Oosterbroek},
  {Oppizzi}, {Ordenovic}, {Osato}, {Pacaud}, {Pace}, {Padilla}, {Paech},
  {Pagano}, {Page}, {Palazzi}, {Paltani}, {Pamuk}, {Pandolfi}, {Paoletti},
  {Paolillo}, {Papaderos}, {Pardede}, {Parimbelli}, {Parmar}, {Partmann},
  {Pasian}, {Passalacqua}, {Paterson}, {Patrizii}, {Pattison},
  {Paulino-Afonso}, {Paviot}, {Peacock}, {Pearce}, {Pedersen}, {Peel},
  {Peletier}, {Pellejero Ibanez}, {Pello}, {Penny}, {Percival},
  {Perez-Garrido}, {Perotto}, {Pettorino}, {Pezzotta}, {Pezzuto}, {Philippon},
  {Pierre}, {Piersanti}, {Pietroni}, {Piga}, {Pilo}, {Pires}, {Pisani},
  {Pizzella}, {Pizzuti}, {Plana}, {Polenta}, {Pollack}, {Poncet},
  {P{\"o}ntinen}, {Pool}, {Popa}, {Popa}, {Popp}, {Porciani}, {Porth},
  {Potter}, {Poulain}, {Pourtsidou}, {Pozzetti}, {Prandoni}, {Pratt},
  {Prezelus}, {Prieto}, {Pugno}, {Quai}, {Quilley}, {Racca}, {Raccanelli},
  {R{\'a}cz}, {Radinovi{\'c}}, {Radovich}, {Ragagnin}, {Ragnit}, {Raison},
  {Ramos-Chernenko}, {Ranc}, {Rasera}, {Raylet}, {Rebolo}, {Refregier},
  {Reimberg}, {Reiprich}, {Renk}, {Renzi}, {Retre}, {Revaz}, {Reyl{\'e}},
  {Reynolds}, {Rhodes}, {Ricci}, {Ricci}, {Riccio}, {Ricken}, {Rissanen},
  {Risso}, {Rix}, {Robin}, {Rocca-Volmerange}, {Rocci}, {Rodenhuis},
  {Rodighiero}, {Rodriguez Monroy}, {Rollins}, {Romanello}, {Roman}, {Romelli},
  {Romero-Gomez}, {Roncarelli}, {Rosati}, {Rosset}, {Rossetti}, {Roster},
  {Rottgering}, {Rozas-Fern{\'a}ndez}, {Ruane}, {Rubino-Martin}, {Rudolph},
  {Ruppin}, {Rusholme}, {Sacquegna}, {S{\'a}ez-Casares}, {Saga}, {Saglia},
  {Sahl{\'e}n}, {Saifollahi}, {Sakr}, {Salvalaggio}, {Salvaterra}, {Salvati},
  {Salvato}, {Salvignol}, {S{\'a}nchez}, {Sanchez}, {Sanders}, {Sapone},
  {Saponara}, {Sarpa}, {Sarron}, {Sartori}, {Sartoris}, {Sassolas}, {Sauniere},
  {Sauvage}, {Sawicki}, {Scaramella}, {Scarlata}, {Scharr{\'e}}, {Schaye},
  {Schewtschenko}, {Schindler}, {Schinnerer}, {Schirmer}, {Schmidt}, {Schmidt},
  {Schmidt}, {Schneider}, {Schneider}, {Schneider}, {Sch{\"o}neberg},
  {Schrabback}, {Schultheis}, {Schulz}, {Schuster}, {Schwartz}, {Sciotti},
  {Scodeggio}, {Scognamiglio}, {Scott}, {Scottez}, {Secroun}, {Sefusatti},
  {Seidel}, {Seiffert}, {Sellentin}, {Selwood}, {Semboloni}, {Sereno},
  {Serjeant}, {Serrano}, {Setnikar}, {Shankar}, {Sharples}, {Short},
  {Shulevski}, {Shuntov}, {Sias}, {Sikkema}, {Silvestri}, {Simon}, {Sirignano},
  {Sirri}, {Skottfelt}, {Slezak}, {Sluse}, {Smith}, {Smith}, {Smith}, {Smit},
  {Soldano}, {Solheim}, {Sorce}, {Sorrenti}, {Soubrie}, {Spinoglio}, {Spurio
  Mancini}, {Stadel}, {Stagnaro}, {Stanco}, {Stanford}, {Starck}, {Stassi},
  {Steinwagner}, {Stern}, {Stone}, {Strada}, {Strafella}, {Stramaccioni},
  {Surace}, {Sureau}, {Suyu}, {Swindells}, {Szafraniec}, {Szapudi}, {Taamoli},
  {Talia}, {Tallada-Cresp{\'\i}}, {Tanidis}, {Tao}, {Tarr{\'\i}o},
  {Tavagnacco}, {Taylor}, {Taylor}, {Taylor}, {Teixeira}, {Tenti}, {Teodoro
  Idiago}, {Teplitz}, {Tereno}, {Tessore}, {Testa}, {Testera}, {Tewes},
  {Teyssier}, {Theret}, {Thizy}, {Thomas}, {Toba}, {Toft}, {Toledo-Moreo},
  {Tolstoy}, {Tommasi}, {Torbaniuk}, {Torradeflot}, {Tortora}, {Tosi}, {Tosti},
  {Trifoglio}, {Troja}, {Trombetti}, {Tronconi}, {Tsedrik}, {Tsyganov},
  {Tucci}, {Tutusaus}, {Uhlemann}, {Ulivi}, {Urbano}, {Vacher}, {Vaillon},
  {Valageas}, {Valdes}, {Valentijn}, {Valenziano}, {Valieri}, {Valiviita}, {Van
  den Broeck}, {Vassallo}, {Vavrek}, {Vega-Ferrero}, {Venemans}, {Venhola},
  {Ventura}, {Verdoes Kleijn}, {Vergani}, {Verma}, {Vernizzi}, {Veropalumbo},
  {Verza}, {Vescovi}, {Vibert}, {Viel}, {Vielzeuf}, {Viglione}, {Viitanen},
  {Villaescusa-Navarro}, {Vinciguerra}, {Visticot}, {Voggel}, {von
  Wietersheim-Kramsta}, {Vriend}, {Wachter}, {Walmsley}, {Walth}, {Walton},
  {Walton}, {Wander}, {Wang}, {Wang}, {Weaver}, {Weller}, {Wetzstein},
  {Whalen}, {Whittam}, {Widmer}, {Wiesmann}, {Wilde}, {Williams}, {Winther},
  {Wittje}, {Wong}, {Wright}, {Yankelevich}, {Yeung}, {Yoon}, {Youles}, {Yung},
  {Zacchei}, {Zalesky}, {Zamorani}, {Zamorano Vitorelli}, {Zanoni Marc},
  {Zennaro}, {Zerbi}, {Zinchenko}, {Zoubian}, {Zucca}, \&
  {Zumalacarregui}}]{EuclidPaper1_2024}
{Euclid Collaboration}, {Mellier}, Y., {Abdurro'uf}, {et~al.} 2024, arXiv
  e-prints, arXiv:2405.13491

\bibitem[{{Foreman-Mackey} {et~al.}(2013){Foreman-Mackey}, {Hogg}, {Lang}, \&
  {Goodman}}]{Emcee2013}
{Foreman-Mackey}, D., {Hogg}, D.~W., {Lang}, D., \& {Goodman}, J. 2013, \pasp,
  125, 306

\bibitem[{{Friedrich} {et~al.}(2021){Friedrich}, {Andrade-Oliveira}, {Camacho},
  {Alves}, {Rosenfeld}, {Sanchez}, {Fang}, {Eifler}, {Krause}, {Chang},
  {Omori}, {Amon}, {Baxter}, {Elvin-Poole}, {Huterer}, {Porredon}, {Prat},
  {Terra}, {Troja}, {Alarcon}, {Bechtol}, {Bernstein}, {Buchs}, {Campos},
  {Carnero Rosell}, {Carrasco Kind}, {Cawthon}, {Choi}, {Cordero}, {Crocce},
  {Davis}, {DeRose}, {Diehl}, {Dodelson}, {Doux}, {Drlica-Wagner}, {Elsner},
  {Everett}, {Fosalba}, {Gatti}, {Giannini}, {Gruen}, {Gruendl}, {Harrison},
  {Hartley}, {Jain}, {Jarvis}, {MacCrann}, {McCullough}, {Muir}, {Myles},
  {Pandey}, {Raveri}, {Roodman}, {Rodriguez-Monroy}, {Rykoff}, {Samuroff},
  {S{\'a}nchez}, {Secco}, {Sevilla-Noarbe}, {Sheldon}, {Troxel}, {Weaverdyck},
  {Yanny}, {Aguena}, {Avila}, {Bacon}, {Bertin}, {Bhargava}, {Brooks}, {Burke},
  {Carretero}, {Costanzi}, {da Costa}, {Pereira}, {De Vicente}, {Desai},
  {Evrard}, {Ferrero}, {Frieman}, {Garc{\'\i}a-Bellido}, {Gaztanaga}, {Gerdes},
  {Giannantonio}, {Gschwend}, {Gutierrez}, {Hinton}, {Hollowood}, {Honscheid},
  {James}, {Kuehn}, {Lahav}, {Lima}, {Maia}, {Menanteau}, {Miquel}, {Morgan},
  {Palmese}, {Paz-Chinch{\'o}n}, {Plazas}, {Sanchez}, {Scarpine}, {Serrano},
  {Soares-Santos}, {Smith}, {Suchyta}, {Tarle}, {Thomas}, {To}, {Varga},
  {Weller}, {Wilkinson}, {Wilkinson}, \& {DES
  Collaboration}}]{Friedrich2021covariance}
{Friedrich}, O., {Andrade-Oliveira}, F., {Camacho}, H., {et~al.} 2021, \mnras,
  508, 3125

\bibitem[{{Gil-Mar{\'\i}n} {et~al.}(2016){Gil-Mar{\'\i}n}, {Percival},
  {Brownstein}, {Chuang}, {Grieb}, {Ho}, {Kitaura}, {Maraston}, {Prada},
  {Rodr{\'\i}guez-Torres}, {Ross}, {Samushia}, {Schlegel}, {Thomas}, {Tinker},
  \& {Zhao}}]{Gil-Marin2016}
{Gil-Mar{\'\i}n}, H., {Percival}, W.~J., {Brownstein}, J.~R., {et~al.} 2016,
  \mnras, 460, 4188

\bibitem[{{Hadzhiyska} {et~al.}(2024){Hadzhiyska}, {Ferraro}, {Ried Guachalla},
  {Schaan}, {Aguilar}, {Battaglia}, {Bond}, {Brooks}, {Calabrese}, {Choi},
  {Claybaugh}, {Coulton}, {Dawson}, {Devlin}, {Dey}, {Doel}, {Duivenvoorden},
  {Dunkley}, {Farren}, {Font-Ribera}, {Forero-Romero}, {Gallardo},
  {Gazta{\~n}aga}, {Gontcho Gontcho}, {Gralla}, {Le Guillou}, {Gutierrez},
  {Guy}, {Hill}, {Hlo{\v{z}}ek}, {Honscheid}, {Juneau}, {Kisner}, {Kremin},
  {Landriau}, {Liu}, {Louis}, {MacCrann}, {de Macorra}, {Madhavacheril},
  {Manera}, {Meisner}, {Miquel}, {Moodley}, {Moustakas}, {Mroczkowski},
  {Naess}, {Newman}, {Niemack}, {Niz}, {Page}, {Palanque-Delabrouille},
  {Partridge}, {Percival}, {Prada}, {Qu}, {Rossi}, {Sanchez}, {Schlegel},
  {Schubnell}, {Sehgal}, {Seo}, {Sif{\'o}n}, {Spergel}, {Sprayberry}, {Staggs},
  {Tarl{\'e}}, {Vargas}, {Vavagiakis}, {Weaver}, {Wollack}, {Zhou}, \&
  {Zou}}]{Hadzhiyska2024}
{Hadzhiyska}, B., {Ferraro}, S., {Ried Guachalla}, B., {et~al.} 2024, arXiv
  e-prints, arXiv:2407.07152

\bibitem[{{Hildebrandt} {et~al.}(2020){Hildebrandt}, {K{\"o}hlinger}, {van den
  Busch}, {Joachimi}, {Heymans}, {Kannawadi}, {Wright}, {Asgari}, {Blake},
  {Hoekstra}, {Joudaki}, {Kuijken}, {Miller}, {Morrison}, {Tr{\"o}ster},
  {Amon}, {Archidiacono}, {Brieden}, {Choi}, {de Jong}, {Erben}, {Giblin},
  {Mead}, {Peacock}, {Radovich}, {Schneider}, {Sif{\'o}n}, \&
  {Tewes}}]{Hildebrandt2020}
{Hildebrandt}, H., {K{\"o}hlinger}, F., {van den Busch}, J.~L., {et~al.} 2020,
  \aap, 633, A69

\bibitem[{{Hildebrandt} {et~al.}(2017){Hildebrandt}, {Viola}, {Heymans},
  {Joudaki}, {Kuijken}, {Blake}, {Erben}, {Joachimi}, {Klaes}, {Miller},
  {Morrison}, {Nakajima}, {Verdoes Kleijn}, {Amon}, {Choi}, {Covone}, {de
  Jong}, {Dvornik}, {Fenech Conti}, {Grado}, {Harnois-D{\'e}raps}, {Herbonnet},
  {Hoekstra}, {K{\"o}hlinger}, {McFarland}, {Mead}, {Merten}, {Napolitano},
  {Peacock}, {Radovich}, {Schneider}, {Simon}, {Valentijn}, {van den Busch},
  {van Uitert}, \& {Van Waerbeke}}]{Hildebrandt2017}
{Hildebrandt}, H., {Viola}, M., {Heymans}, C., {et~al.} 2017, \mnras, 465, 1454

\bibitem[{{Joachimi} {et~al.}(2021){Joachimi}, {Lin}, {Asgari}, {Tr{\"o}ster},
  {Heymans}, {Hildebrandt}, {K{\"o}hlinger}, {S{\'a}nchez}, {Wright},
  {Bilicki}, {Blake}, {van den Busch}, {Crocce}, {Dvornik}, {Erben}, {Getman},
  {Giblin}, {Hoekstra}, {Kannawadi}, {Kuijken}, {Napolitano}, {Schneider},
  {Scoccimarro}, {Sellentin}, {Shan}, {von Wietersheim-Kramsta}, \&
  {Zuntz}}]{Joachimi2021Kids}
{Joachimi}, B., {Lin}, C.~A., {Asgari}, M., {et~al.} 2021, \aap, 646, A129

\bibitem[{{Joachimi} {et~al.}(2011){Joachimi}, {Mandelbaum}, {Abdalla}, \&
  {Bridle}}]{Joachimi2011}
{Joachimi}, B., {Mandelbaum}, R., {Abdalla}, F.~B., \& {Bridle}, S.~L. 2011,
  \aap, 527, A26

\bibitem[{{Joseph} {et~al.}(2023){Joseph}, {Aloni}, {Schmaltz}, {Sivarajan}, \&
  {Weiner}}]{Joseph2023}
{Joseph}, M., {Aloni}, D., {Schmaltz}, M., {Sivarajan}, E.~N., \& {Weiner}, N.
  2023, \prd, 108, 023520

\bibitem[{{Kaiser}(1992)}]{Kaiser1992}
{Kaiser}, N. 1992, \apj, 388, 272

\bibitem[{{Kara{\c{c}}ayl{\i}} {et~al.}(2024){Kara{\c{c}}ayl{\i}}, {Martini},
  {Guy}, {Ravoux}, {Abdul Karim}, {Armengaud}, {Walther}, {Aguilar}, {Ahlen},
  {Bailey}, {Bautista}, {Beltran}, {Brooks}, {Cabayol-Garcia}, {Chabanier},
  {Chaussidon}, {Chaves-Montero}, {Dawson}, {de la Cruz}, {de la Macorra},
  {Doel}, {Font-Ribera}, {Forero-Romero}, {Gontcho}, {Gonzalez-Morales},
  {Gordon}, {Herrera-Alcantar}, {Honscheid}, {Ir{\v{s}}i{\v{c}}}, {Ishak},
  {Kehoe}, {Kisner}, {Kremin}, {Landriau}, {Le Guillou}, {Levi}, {Luki{\'c}},
  {Meisner}, {Miquel}, {Moustakas}, {Mueller}, {Mu{\~n}oz-Guti{\'e}rrez},
  {Napolitano}, {Nie}, {Niz}, {Palanque-Delabrouille}, {Percival}, {Pieri},
  {Poppett}, {Prada}, {P{\'e}rez-R{\`a}fols}, {Ram{\'\i}rez-P{\'e}rez},
  {Rossi}, {Sanchez}, {Seo}, {Sinigaglia}, {Tan}, {Tarl{\'e}}, {Wang},
  {Weaver}, {Y{\'e}che}, \& {Zhou}}]{Karacayli2024}
{Kara{\c{c}}ayl{\i}}, N.~G., {Martini}, P., {Guy}, J., {et~al.} 2024, \mnras,
  528, 3941

\bibitem[{{Kilbinger}(2015)}]{Kilbinger2015}
{Kilbinger}, M. 2015, Reports on Progress in Physics, 78, 086901

\bibitem[{{Krause} {et~al.}(2016){Krause}, {Eifler}, \& {Blazek}}]{Krause2016}
{Krause}, E., {Eifler}, T., \& {Blazek}, J. 2016, \mnras, 456, 207

\bibitem[{{Kuijken} {et~al.}(2019){Kuijken}, {Heymans}, {Dvornik},
  {Hildebrandt}, {de Jong}, {Wright}, {Erben}, {Bilicki}, {Giblin}, {Shan},
  {Getman}, {Grado}, {Hoekstra}, {Miller}, {Napolitano}, {Paolilo}, {Radovich},
  {Schneider}, {Sutherland}, {Tewes}, {Tortora}, {Valentijn}, \& {Verdoes
  Kleijn}}]{Kuijken2019KiDS1000}
{Kuijken}, K., {Heymans}, C., {Dvornik}, A., {et~al.} 2019, \aap, 625, A2

\bibitem[{{Lahav} \& {Liddle}(2022)}]{Lahav2022}
{Lahav}, O. \& {Liddle}, A.~R. 2022, arXiv e-prints, arXiv:2201.08666

\bibitem[{{Li} {et~al.}(2023){Li}, {Zhang}, {Sugiyama}, {Dalal}, {Terasawa},
  {Rau}, {Mandelbaum}, {Takada}, {More}, {Strauss}, {Miyatake}, {Shirasaki},
  {Hamana}, {Oguri}, {Luo}, {Nishizawa}, {Takahashi}, {Nicola}, {Osato},
  {Kannawadi}, {Sunayama}, {Armstrong}, {Bosch}, {Komiyama}, {Lupton}, {Lust},
  {MacArthur}, {Miyazaki}, {Murayama}, {Nishimichi}, {Okura}, {Price}, {Tait},
  {Tanaka}, \& {Wang}}]{Li2023HSCyr3}
{Li}, X., {Zhang}, T., {Sugiyama}, S., {et~al.} 2023, \prd, 108, 123518

\bibitem[{{Limber}(1953)}]{Limber1953approx}
{Limber}, D.~N. 1953, \apj, 117, 134

\bibitem[{{LoVerde} \& {Afshordi}(2008)}]{Loverde2008limber}
{LoVerde}, M. \& {Afshordi}, N. 2008, \prd, 78, 123506

\bibitem[{{LSST Science Collaboration} {et~al.}(2009){LSST Science
  Collaboration}, {Abell}, {Allison}, {Anderson}, {Andrew}, {Angel}, {Armus},
  {Arnett}, {Asztalos}, {Axelrod}, {Bailey}, {Ballantyne}, {Bankert},
  {Barkhouse}, {Barr}, {Barrientos}, {Barth}, {Bartlett}, {Becker}, {Becla},
  {Beers}, {Bernstein}, {Biswas}, {Blanton}, {Bloom}, {Bochanski}, {Boeshaar},
  {Borne}, {Bradac}, {Brandt}, {Bridge}, {Brown}, {Brunner}, {Bullock},
  {Burgasser}, {Burge}, {Burke}, {Cargile}, {Chandrasekharan}, {Chartas},
  {Chesley}, {Chu}, {Cinabro}, {Claire}, {Claver}, {Clowe}, {Connolly}, {Cook},
  {Cooke}, {Cooray}, {Covey}, {Culliton}, {de Jong}, {de Vries}, {Debattista},
  {Delgado}, {Dell'Antonio}, {Dhital}, {Di Stefano}, {Dickinson}, {Dilday},
  {Djorgovski}, {Dobler}, {Donalek}, {Dubois-Felsmann}, {Durech},
  {Eliasdottir}, {Eracleous}, {Eyer}, {Falco}, {Fan}, {Fassnacht}, {Ferguson},
  {Fernandez}, {Fields}, {Finkbeiner}, {Figueroa}, {Fox}, {Francke}, {Frank},
  {Frieman}, {Fromenteau}, {Furqan}, {Galaz}, {Gal-Yam}, {Garnavich},
  {Gawiser}, {Geary}, {Gee}, {Gibson}, {Gilmore}, {Grace}, {Green}, {Gressler},
  {Grillmair}, {Habib}, {Haggerty}, {Hamuy}, {Harris}, {Hawley}, {Heavens},
  {Hebb}, {Henry}, {Hileman}, {Hilton}, {Hoadley}, {Holberg}, {Holman},
  {Howell}, {Infante}, {Ivezic}, {Jacoby}, {Jain}, {R}, {Jedicke}, {Jee},
  {Garrett Jernigan}, {Jha}, {Johnston}, {Jones}, {Juric}, {Kaasalainen},
  {Styliani}, {Kafka}, {Kahn}, {Kaib}, {Kalirai}, {Kantor}, {Kasliwal},
  {Keeton}, {Kessler}, {Knezevic}, {Kowalski}, {Krabbendam}, {Krughoff},
  {Kulkarni}, {Kuhlman}, {Lacy}, {Lepine}, {Liang}, {Lien}, {Lira}, {Long},
  {Lorenz}, {Lotz}, {Lupton}, {Lutz}, {Macri}, {Mahabal}, {Mandelbaum},
  {Marshall}, {May}, {McGehee}, {Meadows}, {Meert}, {Milani}, {Miller},
  {Miller}, {Mills}, {Minniti}, {Monet}, {Mukadam}, {Nakar}, {Neill}, {Newman},
  {Nikolaev}, {Nordby}, {O'Connor}, {Oguri}, {Oliver}, {Olivier}, {Olsen},
  {Olsen}, {Olszewski}, {Oluseyi}, {Padilla}, {Parker}, {Pepper}, {Peterson},
  {Petry}, {Pinto}, {Pizagno}, {Popescu}, {Prsa}, {Radcka}, {Raddick},
  {Rasmussen}, {Rau}, {Rho}, {Rhoads}, {Richards}, {Ridgway}, {Robertson},
  {Roskar}, {Saha}, {Sarajedini}, {Scannapieco}, {Schalk}, {Schindler},
  {Schmidt}, {Schmidt}, {Schneider}, {Schumacher}, {Scranton}, {Sebag},
  {Seppala}, {Shemmer}, {Simon}, {Sivertz}, {Smith}, {Allyn Smith}, {Smith},
  {Spitz}, {Stanford}, {Stassun}, {Strader}, {Strauss}, {Stubbs}, {Sweeney},
  {Szalay}, {Szkody}, {Takada}, {Thorman}, {Trilling}, {Trimble}, {Tyson}, {Van
  Berg}, {Vanden Berk}, {VanderPlas}, {Verde}, {Vrsnak}, {Walkowicz},
  {Wandelt}, {Wang}, {Wang}, {Warner}, {Wechsler}, {West}, {Wiecha},
  {Williams}, {Willman}, {Wittman}, {Wolff}, {Wood-Vasey}, {Wozniak}, {Young},
  {Zentner}, \& {Zhan}}]{abell2009lsst}
{LSST Science Collaboration}, {Abell}, P.~A., {Allison}, J., {et~al.} 2009,
  arXiv e-prints, arXiv:0912.0201

\bibitem[{{Lucca}(2021)}]{Lucca2021}
{Lucca}, M. 2021, Physics of the Dark Universe, 34, 100899

\bibitem[{{McCarthy} {et~al.}(2017){McCarthy}, {Schaye}, {Bird}, \& {Le
  Brun}}]{mcchartybahamas2017}
{McCarthy}, I.~G., {Schaye}, J., {Bird}, S., \& {Le Brun}, A. M.~C. 2017,
  \mnras, 465, 2936

\bibitem[{{Mead} {et~al.}(2015){Mead}, {Peacock}, {Heymans}, {Joudaki}, \&
  {Heavens}}]{Mead2015}
{Mead}, A.~J., {Peacock}, J.~A., {Heymans}, C., {Joudaki}, S., \& {Heavens},
  A.~F. 2015, \mnras, 454, 1958

\bibitem[{{Myles} {et~al.}(2021){Myles}, {Alarcon}, {Amon}, {S{\'a}nchez},
  {Everett}, {DeRose}, {McCullough}, {Gruen}, {Bernstein}, {Troxel},
  {Dodelson}, {Campos}, {MacCrann}, {Yin}, {Raveri}, {Amara}, {Becker}, {Choi},
  {Cordero}, {Eckert}, {Gatti}, {Giannini}, {Gschwend}, {Gruendl}, {Harrison},
  {Hartley}, {Huff}, {Kuropatkin}, {Lin}, {Masters}, {Miquel}, {Prat},
  {Roodman}, {Rykoff}, {Sevilla-Noarbe}, {Sheldon}, {Wechsler}, {Yanny},
  {Abbott}, {Aguena}, {Allam}, {Annis}, {Bacon}, {Bertin}, {Bhargava},
  {Bridle}, {Brooks}, {Burke}, {Carnero Rosell}, {Carrasco Kind}, {Carretero},
  {Castander}, {Conselice}, {Costanzi}, {Crocce}, {da Costa}, {Pereira},
  {Desai}, {Diehl}, {Eifler}, {Elvin-Poole}, {Evrard}, {Ferrero}, {Fert{\'e}},
  {Flaugher}, {Fosalba}, {Frieman}, {Garc{\'\i}a-Bellido}, {Gaztanaga},
  {Giannantonio}, {Hinton}, {Hollowood}, {Honscheid}, {Hoyle}, {Huterer},
  {James}, {Krause}, {Kuehn}, {Lahav}, {Lima}, {Maia}, {Marshall}, {Martini},
  {Melchior}, {Menanteau}, {Mohr}, {Morgan}, {Muir}, {Ogando}, {Palmese},
  {Paz-Chinch{\'o}n}, {Plazas}, {Rodriguez-Monroy}, {Samuroff}, {Sanchez},
  {Scarpine}, {Secco}, {Serrano}, {Smith}, {Soares-Santos}, {Suchyta},
  {Swanson}, {Tarle}, {Thomas}, {To}, {Varga}, {Weller}, \&
  {Wester}}]{Milesetal2021DESphotz}
{Myles}, J., {Alarcon}, A., {Amon}, A., {et~al.} 2021, \mnras, 505, 4249

\bibitem[{{Pakmor} {et~al.}(2023){Pakmor}, {Springel}, {Coles}, {Guillet},
  {Pfrommer}, {Bose}, {Barrera}, {Delgado}, {Ferlito}, {Frenk}, {Hadzhiyska},
  {Hern{\'a}ndez-Aguayo}, {Hernquist}, {Kannan}, \& {White}}]{Pakmor2023}
{Pakmor}, R., {Springel}, V., {Coles}, J.~P., {et~al.} 2023, \mnras, 524, 2539

\bibitem[{{Palanque-Delabrouille} {et~al.}(2015){Palanque-Delabrouille},
  {Y{\`e}che}, {Baur}, {Magneville}, {Rossi}, {Lesgourgues}, {Borde}, {Burtin},
  {LeGoff}, {Rich}, {Viel}, \& {Weinberg}}]{Palanque2015}
{Palanque-Delabrouille}, N., {Y{\`e}che}, C., {Baur}, J., {et~al.} 2015, \jcap,
  2015, 011

\bibitem[{{Planck Collaboration} {et~al.}(2020{\natexlab{a}}){Planck
  Collaboration}, {Aghanim}, {Akrami}, {Ashdown}, {Aumont}, {Baccigalupi},
  {Ballardini}, {Banday}, {Barreiro}, {Bartolo}, {Basak}, {Battye}, {Benabed},
  {Bernard}, {Bersanelli}, {Bielewicz}, {Bock}, {Bond}, {Borrill}, {Bouchet},
  {Boulanger}, {Bucher}, {Burigana}, {Butler}, {Calabrese}, {Cardoso},
  {Carron}, {Challinor}, {Chiang}, {Chluba}, {Colombo}, {Combet}, {Contreras},
  {Crill}, {Cuttaia}, {de Bernardis}, {de Zotti}, {Delabrouille}, {Delouis},
  {Di Valentino}, {Diego}, {Dor{\'e}}, {Douspis}, {Ducout}, {Dupac}, {Dusini},
  {Efstathiou}, {Elsner}, {En{\ss}lin}, {Eriksen}, {Fantaye}, {Farhang},
  {Fergusson}, {Fernandez-Cobos}, {Finelli}, {Forastieri}, {Frailis},
  {Fraisse}, {Franceschi}, {Frolov}, {Galeotta}, {Galli}, {Ganga},
  {G{\'e}nova-Santos}, {Gerbino}, {Ghosh}, {Gonz{\'a}lez-Nuevo}, {G{\'o}rski},
  {Gratton}, {Gruppuso}, {Gudmundsson}, {Hamann}, {Handley}, {Hansen},
  {Herranz}, {Hildebrandt}, {Hivon}, {Huang}, {Jaffe}, {Jones}, {Karakci},
  {Keih{\"a}nen}, {Keskitalo}, {Kiiveri}, {Kim}, {Kisner}, {Knox},
  {Krachmalnicoff}, {Kunz}, {Kurki-Suonio}, {Lagache}, {Lamarre}, {Lasenby},
  {Lattanzi}, {Lawrence}, {Le Jeune}, {Lemos}, {Lesgourgues}, {Levrier},
  {Lewis}, {Liguori}, {Lilje}, {Lilley}, {Lindholm}, {L{\'o}pez-Caniego},
  {Lubin}, {Ma}, {Mac{\'\i}as-P{\'e}rez}, {Maggio}, {Maino}, {Mandolesi},
  {Mangilli}, {Marcos-Caballero}, {Maris}, {Martin}, {Martinelli},
  {Mart{\'\i}nez-Gonz{\'a}lez}, {Matarrese}, {Mauri}, {McEwen}, {Meinhold},
  {Melchiorri}, {Mennella}, {Migliaccio}, {Millea}, {Mitra},
  {Miville-Desch{\^e}nes}, {Molinari}, {Montier}, {Morgante}, {Moss}, {Natoli},
  {N{\o}rgaard-Nielsen}, {Pagano}, {Paoletti}, {Partridge}, {Patanchon},
  {Peiris}, {Perrotta}, {Pettorino}, {Piacentini}, {Polastri}, {Polenta},
  {Puget}, {Rachen}, {Reinecke}, {Remazeilles}, {Renzi}, {Rocha}, {Rosset},
  {Roudier}, {Rubi{\~n}o-Mart{\'\i}n}, {Ruiz-Granados}, {Salvati}, {Sandri},
  {Savelainen}, {Scott}, {Shellard}, {Sirignano}, {Sirri}, {Spencer},
  {Sunyaev}, {Suur-Uski}, {Tauber}, {Tavagnacco}, {Tenti}, {Toffolatti},
  {Tomasi}, {Trombetti}, {Valenziano}, {Valiviita}, {Van Tent}, {Vibert},
  {Vielva}, {Villa}, {Vittorio}, {Wandelt}, {Wehus}, {White}, {White},
  {Zacchei}, \& {Zonca}}]{PlanckVI2020}
{Planck Collaboration}, {Aghanim}, N., {Akrami}, Y., {et~al.}
  2020{\natexlab{a}}, \aap, 641, A6

\bibitem[{{Planck Collaboration} {et~al.}(2020{\natexlab{b}}){Planck
  Collaboration}, {Aghanim}, {Akrami}, {Ashdown}, {Aumont}, {Baccigalupi},
  {Ballardini}, {Banday}, {Barreiro}, {Bartolo}, {Basak}, {Benabed}, {Bernard},
  {Bersanelli}, {Bielewicz}, {Bock}, {Bond}, {Borrill}, {Bouchet}, {Boulanger},
  {Bucher}, {Burigana}, {Calabrese}, {Cardoso}, {Carron}, {Challinor},
  {Chiang}, {Colombo}, {Combet}, {Crill}, {Cuttaia}, {de Bernardis}, {de
  Zotti}, {Delabrouille}, {Di Valentino}, {Diego}, {Dor{\'e}}, {Douspis},
  {Ducout}, {Dupac}, {Efstathiou}, {Elsner}, {En{\ss}lin}, {Eriksen},
  {Fantaye}, {Fernandez-Cobos}, {Finelli}, {Forastieri}, {Frailis}, {Fraisse},
  {Franceschi}, {Frolov}, {Galeotta}, {Galli}, {Ganga}, {G{\'e}nova-Santos},
  {Gerbino}, {Ghosh}, {Gonz{\'a}lez-Nuevo}, {G{\'o}rski}, {Gratton},
  {Gruppuso}, {Gudmundsson}, {Hamann}, {Handley}, {Hansen}, {Herranz}, {Hivon},
  {Huang}, {Jaffe}, {Jones}, {Karakci}, {Keih{\"a}nen}, {Keskitalo}, {Kiiveri},
  {Kim}, {Knox}, {Krachmalnicoff}, {Kunz}, {Kurki-Suonio}, {Lagache},
  {Lamarre}, {Lasenby}, {Lattanzi}, {Lawrence}, {Le Jeune}, {Levrier}, {Lewis},
  {Liguori}, {Lilje}, {Lindholm}, {L{\'o}pez-Caniego}, {Lubin}, {Ma},
  {Mac{\'\i}as-P{\'e}rez}, {Maggio}, {Maino}, {Mandolesi}, {Mangilli},
  {Marcos-Caballero}, {Maris}, {Martin}, {Mart{\'\i}nez-Gonz{\'a}lez},
  {Matarrese}, {Mauri}, {McEwen}, {Melchiorri}, {Mennella}, {Migliaccio},
  {Miville-Desch{\^e}nes}, {Molinari}, {Moneti}, {Montier}, {Morgante}, {Moss},
  {Natoli}, {Pagano}, {Paoletti}, {Partridge}, {Patanchon}, {Perrotta},
  {Pettorino}, {Piacentini}, {Polastri}, {Polenta}, {Puget}, {Rachen},
  {Reinecke}, {Remazeilles}, {Renzi}, {Rocha}, {Rosset}, {Roudier},
  {Rubi{\~n}o-Mart{\'\i}n}, {Ruiz-Granados}, {Salvati}, {Sandri}, {Savelainen},
  {Scott}, {Sirignano}, {Sunyaev}, {Suur-Uski}, {Tauber}, {Tavagnacco},
  {Tenti}, {Toffolatti}, {Tomasi}, {Trombetti}, {Valiviita}, {Van Tent},
  {Vielva}, {Villa}, {Vittorio}, {Wandelt}, {Wehus}, {White}, {White},
  {Zacchei}, \& {Zonca}}]{Planck2020CMBlensing}
{Planck Collaboration}, {Aghanim}, N., {Akrami}, Y., {et~al.}
  2020{\natexlab{b}}, \aap, 641, A8

\bibitem[{{Preston} {et~al.}(2023){Preston}, {Amon}, \&
  {Efstathiou}}]{Preston2023_DESyr3}
{Preston}, C., {Amon}, A., \& {Efstathiou}, G. 2023, \mnras, 525, 5554

\bibitem[{{Preston} {et~al.}(2024){Preston}, {Amon}, \&
  {Efstathiou}}]{Preston2024_future}
{Preston}, C., {Amon}, A., \& {Efstathiou}, G. 2024, \mnras, 533, 621

\bibitem[{{Rau} {et~al.}(2023){Rau}, {Dalal}, {Zhang}, {Li}, {Nishizawa},
  {More}, {Mandelbaum}, {Miyatake}, {Strauss}, \& {Takada}}]{Rau2024HSCphotoz}
{Rau}, M.~M., {Dalal}, R., {Zhang}, T., {et~al.} 2023, \mnras, 524, 5109

\bibitem[{{Reid} {et~al.}(2010){Reid}, {Percival}, {Eisenstein}, {Verde},
  {Spergel}, {Skibba}, {Bahcall}, {Budavari}, {Frieman}, {Fukugita}, {Gott},
  {Gunn}, {Ivezi{\'c}}, {Knapp}, {Kron}, {Lupton}, {McKay}, {Meiksin},
  {Nichol}, {Pope}, {Schlegel}, {Schneider}, {Stoughton}, {Strauss}, {Szalay},
  {Tegmark}, {Vogeley}, {Weinberg}, {York}, \& {Zehavi}}]{Reid2010}
{Reid}, B.~A., {Percival}, W.~J., {Eisenstein}, D.~J., {et~al.} 2010, \mnras,
  404, 60

\bibitem[{{Riess} {et~al.}(2021){Riess}, {Casertano}, {Yuan}, {Bowers},
  {Macri}, {Zinn}, \& {Scolnic}}]{Riess2021}
{Riess}, A.~G., {Casertano}, S., {Yuan}, W., {et~al.} 2021, \apjl, 908, L6

\bibitem[{{Schaller} {et~al.}(2024){Schaller}, {Schaye}, {Kugel}, {Broxterman},
  \& {van Daalen}}]{Schaller2024}
{Schaller}, M., {Schaye}, J., {Kugel}, R., {Broxterman}, J.~C., \& {van
  Daalen}, M.~P. 2024, arXiv e-prints, arXiv:2410.17109

\bibitem[{{Schaye} {et~al.}(2023){Schaye}, {Kugel}, {Schaller}, {Helly},
  {Braspenning}, {Elbers}, {McCarthy}, {van Daalen}, {Vandenbroucke}, {Frenk},
  {Kwan}, {Salcido}, {Bah{\'e}}, {Borrow}, {Chaikin}, {Hahn}, {Hu{\v{s}}ko},
  {Jenkins}, {Lacey}, \& {Nobels}}]{Schaye2023}
{Schaye}, J., {Kugel}, R., {Schaller}, M., {et~al.} 2023, \mnras, 526, 4978

\bibitem[{{Schneider} {et~al.}(2010){Schneider}, {Eifler}, \&
  {Krause}}]{Schneider2010cosebis}
{Schneider}, P., {Eifler}, T., \& {Krause}, E. 2010, \aap, 520, A116

\bibitem[{{Schneider} {et~al.}(2002){Schneider}, {van Waerbeke}, {Kilbinger},
  \& {Mellier}}]{Schneider2002}
{Schneider}, P., {van Waerbeke}, L., {Kilbinger}, M., \& {Mellier}, Y. 2002,
  \aap, 396, 1

\bibitem[{{Secco} {et~al.}(2022){Secco}, {Samuroff}, {Krause}, {Jain},
  {Blazek}, {Raveri}, {Campos}, {Amon}, {Chen}, {Doux}, {Choi}, {Gruen},
  {Bernstein}, {Chang}, {DeRose}, {Myles}, {Fert{\'e}}, {Lemos}, {Huterer},
  {Prat}, {Troxel}, {MacCrann}, {Liddle}, {Kacprzak}, {Fang}, {S{\'a}nchez},
  {Pandey}, {Dodelson}, {Chintalapati}, {Hoffmann}, {Alarcon}, {Alves},
  {Andrade-Oliveira}, {Baxter}, {Bechtol}, {Becker}, {Brandao-Souza},
  {Camacho}, {Carnero Rosell}, {Carrasco Kind}, {Cawthon}, {Cordero}, {Crocce},
  {Davis}, {Di Valentino}, {Drlica-Wagner}, {Eckert}, {Eifler}, {Elidaiana},
  {Elsner}, {Elvin-Poole}, {Everett}, {Fosalba}, {Friedrich}, {Gatti},
  {Giannini}, {Gruendl}, {Harrison}, {Hartley}, {Herner}, {Huang}, {Huff},
  {Jarvis}, {Jeffrey}, {Kuropatkin}, {Leget}, {Muir}, {Mccullough}, {Navarro
  Alsina}, {Omori}, {Park}, {Porredon}, {Rollins}, {Roodman}, {Rosenfeld},
  {Ross}, {Rykoff}, {Sanchez}, {Sevilla-Noarbe}, {Sheldon}, {Shin}, {Troja},
  {Tutusaus}, {Varga}, {Weaverdyck}, {Wechsler}, {Yanny}, {Yin}, {Zhang},
  {Zuntz}, {Abbott}, {Aguena}, {Allam}, {Annis}, {Bacon}, {Bertin}, {Bhargava},
  {Bridle}, {Brooks}, {Buckley-Geer}, {Burke}, {Carretero}, {Costanzi}, {da
  Costa}, {De Vicente}, {Diehl}, {Dietrich}, {Doel}, {Ferrero}, {Flaugher},
  {Frieman}, {Garc{\'\i}a-Bellido}, {Gaztanaga}, {Gerdes}, {Giannantonio},
  {Gschwend}, {Gutierrez}, {Hinton}, {Hollowood}, {Honscheid}, {Hoyle},
  {James}, {Jeltema}, {Kuehn}, {Lahav}, {Lima}, {Lin}, {Maia}, {Marshall},
  {Martini}, {Melchior}, {Menanteau}, {Miquel}, {Mohr}, {Morgan}, {Ogando},
  {Palmese}, {Paz-Chinch{\'o}n}, {Petravick}, {Pieres}, {Plazas Malag{\'o}n},
  {Rodriguez-Monroy}, {Romer}, {Sanchez}, {Scarpine}, {Schubnell}, {Scolnic},
  {Serrano}, {Smith}, {Soares-Santos}, {Suchyta}, {Swanson}, {Tarle}, {Thomas},
  {To}, \& {DES Collaboration}}]{Secco2022}
{Secco}, L.~F., {Samuroff}, S., {Krause}, E., {et~al.} 2022, \prd, 105, 023515

\bibitem[{{Shirasaki} {et~al.}(2019){Shirasaki}, {Hamana}, {Takada},
  {Takahashi}, \& {Miyatake}}]{Shirasaki19HSCCov}
{Shirasaki}, M., {Hamana}, T., {Takada}, M., {Takahashi}, R., \& {Miyatake}, H.
  2019, \mnras, 486, 52

\bibitem[{{Spergel} {et~al.}(2015){Spergel}, {Gehrels}, {Baltay}, {Bennett},
  {Breckinridge}, {Donahue}, {Dressler}, {Gaudi}, {Greene}, {Guyon}, {Hirata},
  {Kalirai}, {Kasdin}, {Macintosh}, {Moos}, {Perlmutter}, {Postman},
  {Rauscher}, {Rhodes}, {Wang}, {Weinberg}, {Benford}, {Hudson}, {Jeong},
  {Mellier}, {Traub}, {Yamada}, {Capak}, {Colbert}, {Masters}, {Penny},
  {Savransky}, {Stern}, {Zimmerman}, {Barry}, {Bartusek}, {Carpenter}, {Cheng},
  {Content}, {Dekens}, {Demers}, {Grady}, {Jackson}, {Kuan}, {Kruk}, {Melton},
  {Nemati}, {Parvin}, {Poberezhskiy}, {Peddie}, {Ruffa}, {Wallace}, {Whipple},
  {Wollack}, \& {Zhao}}]{spergel2015wide}
{Spergel}, D., {Gehrels}, N., {Baltay}, C., {et~al.} 2015, arXiv e-prints,
  arXiv:1503.03757

\bibitem[{{Tanimura} {et~al.}(2023){Tanimura}, {Douspis}, {Aghanim}, \&
  {Kuruvilla}}]{Tanimura2023}
{Tanimura}, H., {Douspis}, M., {Aghanim}, N., \& {Kuruvilla}, J. 2023, \aap,
  674, A222

\bibitem[{{Tegmark} {et~al.}(2004){Tegmark}, {Strauss}, {Blanton}, {Abazajian},
  {Dodelson}, {Sandvik}, {Wang}, {Weinberg}, {Zehavi}, {Bahcall}, {Hoyle},
  {Schlegel}, {Scoccimarro}, {Vogeley}, {Berlind}, {Budavari}, {Connolly},
  {Eisenstein}, {Finkbeiner}, {Frieman}, {Gunn}, {Hui}, {Jain}, {Johnston},
  {Kent}, {Lin}, {Nakajima}, {Nichol}, {Ostriker}, {Pope}, {Scranton},
  {Seljak}, {Sheth}, {Stebbins}, {Szalay}, {Szapudi}, {Xu}, {Annis},
  {Brinkmann}, {Burles}, {Castander}, {Csabai}, {Loveday}, {Doi}, {Fukugita},
  {Gillespie}, {Hennessy}, {Hogg}, {Ivezi{\'c}}, {Knapp}, {Lamb}, {Lee},
  {Lupton}, {McKay}, {Kunszt}, {Munn}, {O'Connell}, {Peoples}, {Pier},
  {Richmond}, {Rockosi}, {Schneider}, {Stoughton}, {Tucker}, {vanden Berk},
  {Yanny}, \& {York}}]{Tegmark2004}
{Tegmark}, M., {Strauss}, M.~A., {Blanton}, M.~R., {et~al.} 2004, \prd, 69,
  103501

\end{thebibliography}
%

\begin{appendix}
\onecolumn
 \section{KiDS-1000 COSEBI prediction}\label{app:cosebi_measurement}
\begin{figure*}[h]
    \centering
    \includegraphics[width=17cm]{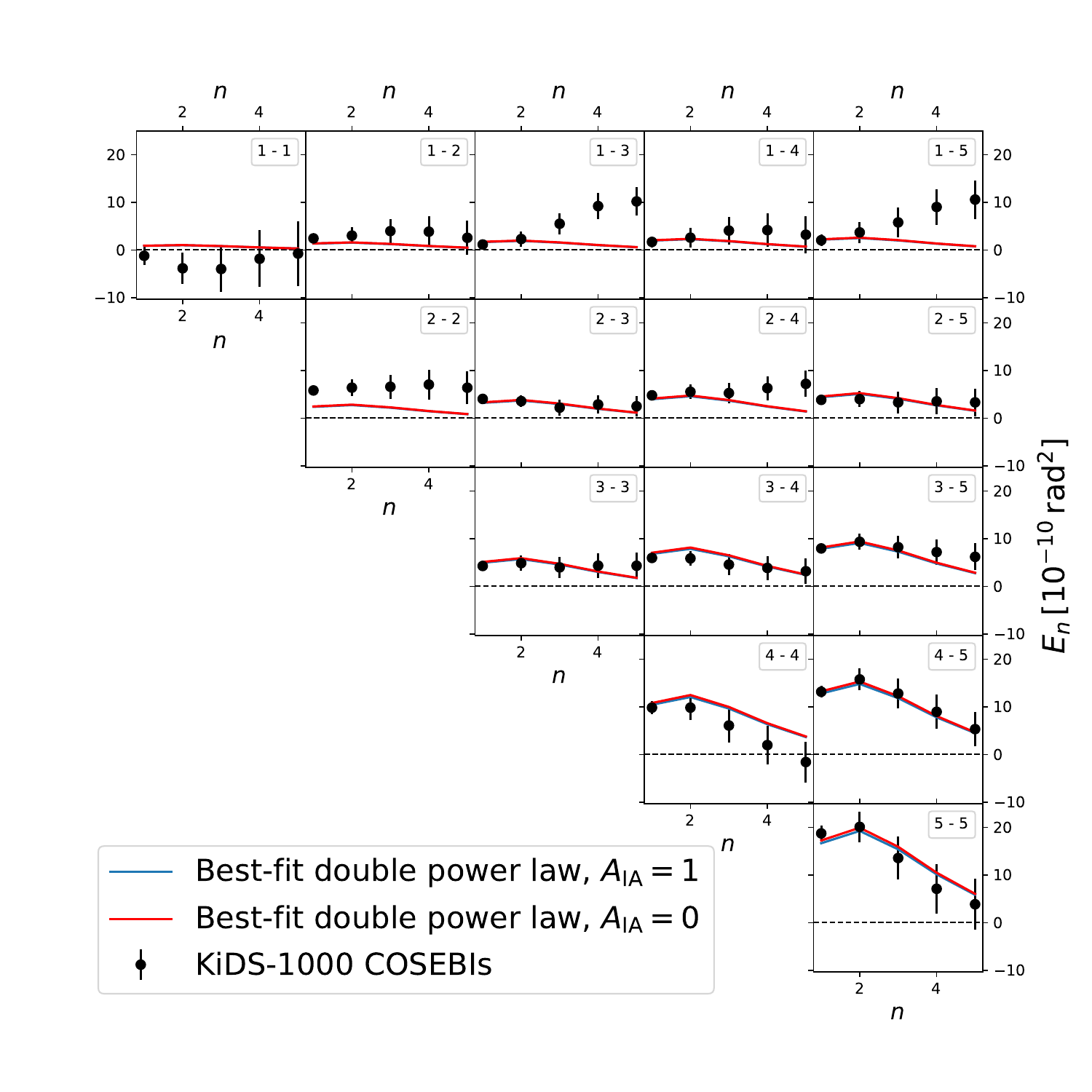}
      \caption{KiDS-1000 COSEBIs measurement (black scatter points) with best-fitting $A_\mathrm{IA}$ = 1 and 0 double power law (solid blue and curves, respectively) for the first five COSEBIs $n$-modes. Each subpanel corresponds to a combination of two tomographic bins, as indicated in the upper right corner.}
         \label{fig:measurments}
    \end{figure*}

            This Appendix shows the best-fitting prediction for the double power-law model to the KiDS-1000 COSEBIs measurements. The scatter points in Fig.~\ref{fig:measurments} are the KiDS-1000 measurements \citep[][]{Asgari2021KiDS}, and the red and blue curves are the best-fitting double power-law fits with $A_\mathrm{IA} = 0$ and $A_\mathrm{IA} = 1$, respectively. The subpanels show the measured $E_n$ as a function of the COSEBI $n$-modes for the combination of two tomographic bins, as indicated in the upper-right corner of each panel. There is no difference in the fitting accuracy with or without including IA, as seen from the reduced chi-square values in Table~\ref{tab:pivot_points}. The double power-law estimate fits the same ranges well as the best-fitting $\Lambda$CDM prediction from \citet{Asgari2021KiDS}. For example, the $\Lambda$CDM and our double power-law estimate both provide good fits to most of the data, but they underpredict the signal for the high-$n$ modes in the ($1-5$) tomographic bin combination and across the entire range for the ($2-2$) combination. 

\end{appendix}

\end{document}